\documentclass[a4paper,12pt]{article}
\usepackage[letter paper, margin=1 in]{geometry}
\usepackage{authblk}
\usepackage{natbib}
\usepackage[utf8]{inputenc}
\usepackage{graphicx}
\usepackage{amsmath}
\usepackage[version=4]{mhchem}
\usepackage{siunitx}
\usepackage{longtable,tabularx}
\usepackage{microtype}
\usepackage{subcaption}
\usepackage{booktabs} 
\usepackage{float}
\usepackage[algo2e,ruled,vlined,linesnumbered]{algorithm2e}
\usepackage{amsfonts}
\usepackage{bm,bbm}
\usepackage{cleveref}
\usepackage[algo2e,ruled,vlined,linesnumbered]{algorithm2e}
\usepackage{url}
\newcommand{\mbb}{\mathbb}
\newcommand{\mbf}{\mathbf}
\newcommand{\mcl}{\mathcal}
\newcommand{\bs}{\boldsymbol}

\newcommand{\f}{\frac}

\newcommand{\T}{\textnormal}
\newcommand{\x}{\mathbf{x}}
\newcommand{\y}{\mathbf{y}}
\newcommand{\X}{\bm{\mathcal{X}}}
\newcommand{\D}{\bm{\mathcal{D}}}
\newcommand{\Sc}{\bm{\mathcal{S}}}
\newcommand{\mfcv}{\texttt{MFCV}}

\newcommand{\defeq}{:=}

\usepackage[normalem]{ulem} 
\newcommand{\argmax}{\operatornamewithlimits{arg\,max}}

\title{Multifidelity Cross-validation}
\author[1,2]{Ashwin Renganathan\footnote{Assistant professor, Aerospace Engineering and the Institute of Computational and Data Science (ICDS), 229 Hammond Building; and AIAA Member.}}
\author[1,2]{Kade Carlson\footnote{Graduate research assistant, Aerospace Engineering, 229 Hammond Building; and AIAA Student Member.}}
\affil[1]{The Pennsylvania State University, University Park, PA, 16802}
\affil[t]{Penn State Institute of Computational and Data Sciences, University Park, PA, 16802}

\begin{document}
\date{}
\maketitle

\begin{abstract}
Emulating the mapping between quantities of interest and their control parameters using surrogate models finds widespread application in engineering design, including in numerical optimization and uncertainty quantification. Gaussian process models can serve as a probabilistic surrogate model of unknown functions, thereby making them highly suitable for engineering design and decision-making in the presence of uncertainty. In this work, we are interested in emulating quantities of interest observed from models of a system at multiple fidelities, which trade accuracy for computational efficiency. Using multifidelity Gaussian process models, to efficiently fuse models at multiple fidelities, we propose a novel method to actively learn the surrogate model via leave-one-out cross-validation (LOO-CV). Our proposed multifidelity cross-validation (\texttt{MFCV}) approach develops an adaptive approach to reduce the LOO-CV error at the target (highest) fidelity, by learning the correlations between the LOO-CV at all fidelities. \texttt{MFCV} develops a two-step lookahead policy to select optimal input-fidelity pairs, both in sequence and in batches, both for continuous and discrete fidelity spaces. We demonstrate the utility of our method on several synthetic test problems as well as on the thermal stress analysis of a gas turbine blade.
\end{abstract}




\section{Introduction}
Historically, the design, certification, and qualification of aircraft and aircraft subsystems have depended heavily on wind tunnel and flight tests. These are risky, expensive, prone to scheduling delays, and can depend on uncontrollable weather conditions. Aircraft design is increasingly benefiting from advances in computational structural and fluid dynamics, and high-performance computing, to minimize the amount of physical experimentation and testing necessary. On the certification front, flight and engine certification by analysis (CbA) is a new paradigm in which airworthiness certificates for aircraft can be obtained via calculations with quantifiable accuracy \cite{mauery202220} \footnote{see \S 25.21 of Chapter 1, Title 14 in the electronic Code for Regulations (CFR) of the Federal Aviation Administrations (FAA).}. 
While this opens up possibilities for significant reduction in the need for physical experimentation in aircraft design and certification, replacing them with computation, several challenges remain. First, computations of high physics fidelity, while being cheaper than wind-tunnel/flight tests, still incur a non-trivial cost. For instance, a ``full fidelity'' direct numerical Navier-Stokes simulation (DNS) of an aircraft under realistic operating conditions is still not practical owing to an intractable number of degrees of freedom in the discretized partial differential equation (PDE) system. Large eddy simulations (LES), while cheaper than DNS, still can cost in the order of a few days of wall-clock time, for full aircraft configurations. Reynolds-averaged Navier-Stokes (RANS) simulations cost several hours of wall-clock time while being unreliable under extreme boundary conditions (e.g., high-angles of attack and side-slip angles). Second, to truly exploit computations for reliable decision-making, all significant forms of uncertainty must be rigorously quantified. This necessitates several thousands of evaluations of the high-fidelity model; even with cheaper RANS models, this can be prohibitive. To alleviate these challenges, surrogate models that trade an acceptable measure of accuracy for significantly large gains in computational efficiency are necessary. Furthermore, surrogate models that can fuse information from models at multiple levels of fidelity (e.g., DNS/LES/RANS) are necessary. This forms the overarching goal of this work.

We exploit the availability of models with tunable fidelities and propose multifidelity methods~\cite{takeno2019multi, kandasamy2017multi, klein2017fast, swersky2013multi} which can judiciously utilize the cost-accuracy trade-off in models to adaptively construct accurate surrogate models. Surrogate models for conservation laws can be efficiently constructed via reduced-order models (ROMs)~\cite{renganathan2018koopman, renganathan2020koopman, renganathan2020machine,renganathan2018methodology} for state variables and general regression/interpolation techniques~\cite{renganathan2021lookahead, renganathan2021enhanced,ashwin2022data, renganathan2023camera,renganathan2021data} for scalar quantities of interest (QoI).
In this work, we are interested in adaptively constructing a data-driven surrogate model for a scalar-valued QoI in order to conservatively use the computational resources. 
 For this purpose, the computational cost of querying models at each fidelity level is built in to the overall methodology. We will leverage a Gaussian process (GP)~\cite{rasmussen:williams:2006} regression model framework to learn the mapping between the QoIs and the inputs. 

The literature covering the use of GP regression for expensive multidisciplinary design of aerospace systems is very broad. This includes aerodynamic design optimization~\cite{renganathan2021enhanced}, uncertainty quantification/reliability analysis~\cite{ranjan2008sequential, bichon2008efficient, picheny2010adaptive, bect2012sequential,echard2010kriging,gotovos2013active,dubourg2013metamodel,wang2016gaussian,marques2018contour,cole2021entropy,booth2023contour,booth2024actively}, and general global optimization~\cite{jones1998efficient,jones2001taxonomy,renganathan2023qpots} just to name a few. This also includes addressing the multifidelity setting as we do. The most popular approach for multifidelity analysis and design using GP regression is that of~\citet{kennedy2001bayesian}, which learns a linear correlation between each fidelity level and a level lower; this results in an autoregressive model that can predict the output of a higher fidelity model from the output of an immediate lower fidelity model. Similar ideas have been explored by \citet{poloczek2017multi}, and further extended to nonlinear correlations by~\citet{perdikaris2017nonlinear}. As opposed to the aforementioned existing work, we propose the use of a GP for the multifidelity setting, that keeps the structure and model inference consistent, and simple, as in the single fidelity setting---more details will follow in \Cref{sec:mfgp}. The next ingredient of our proposed work is active learning. Active learning of GP models refers to the adaptive and automated selection of the training points in a goal-oriented fashion until a relevant convergence criterion is met. 
Typically, a utility function is defined that quantifies the impact of a candidate point on the QoI, and training data are selected by solving an ``inner'' optimization problem. The popular approaches include maximizing the mean-squared error (MSE)~\cite{jin2002sequential,martin2002use} or the integrated MSE (IMSE)~\cite{santner2003design,picheny2010adaptive}, information entropy~\cite{shewry1987maximum,koehler19969}, and mutual information~\cite{krause2008near,beck2016sequential}. See \cite{garud2017design} and \cite{liu2018survey} for a survey of adaptive methods.

Cross-validation (CV)~\cite{rasmussen:williams:2006} refers to quantifying model error via a held out testing set, and could be considered another criterion for active learning. When one training point is held out for testing, and this choice is permuted over all of the training set, this is called leave-one-out cross-validation (LOO-CV)~\cite{li2010accumulative,le2015cokriging,aute2013cross,liu2016adaptive}. Generally, in the context of GPs, CV and/or LOO-CV are used to estimate the hyperparameters of the GP itself~\cite{bachoc2013cross}. However, in this work, we seek to exploit the GP CV error to develop an adaptive sampling strategy for global fit. Recently, \citet{mohammadi2022cross} proposed an adaptive LOO-CV technique where a second GP is fit over the CV errors, which is  in turn used in a Bayesian adaptive setting such as in \cite{jones1998efficient} for active learning. By estimating the CV error at each of the training points, which assumes a non-central chi-squared distribution, they use the expected value of this distribution at each of the training points to fit another GP.
In this work, we take a similar route, but we differ significantly in two fundamental ways. First, we fit a GP to the logarithm of the CV error to comply with the general infinite support of Gaussian random variables. While \cite{mohammadi2022cross} ignores the uncertainty in the CV error, our method can seamlessly include that using heteroscedastic GPs~\cite{binois2018practical,le2005heteroscedastic,ankenman2008stochastic,ankenman2010stochastic}.
Second, we extend this approach to the multifidelity setting where we present a generic acquisition function that can be easily customized to generate novel acquisition functions. Third, our method is \emph{cost-aware}, thereby making sequential decisions that accounts for the cost of querying each specific level.
Finally, our proposed approach seamlessly extends to discrete and continuous levels of fidelity, and sequential as well as batch acquisition settings. In summary, we propose multifidelity cross-validation (\mfcv), which presents a generic method to actively learn a surrogate model from multiple information sources, with the aforementioned features. Our approach is endowed with a straightforward implementation without any necessary user-defined fine tuning. We demonstrate our method on several synthetic test functions as well as real-world experiments. 

The remainder of the article is organized as follows. In \Cref{sec:method}, we provide the background and preliminaries including details about multifidelity GP models. In \Cref{sec:proposed_method}, we provide the details of our method. In \Cref{sec:num_expts}, we show results of our method demonstrated on the synthetic test functions and a real-world stress analysis of a gas turbine blade,
and provide concluding remarks in \Cref{sec:conclusion}.

\section{Background and preliminaries}
\label{sec:method}
\subsection{Gaussian process regression}
\label{sec:gp}
Let $\x \in \X \subset \mbb{R}^d$ be the input to the function $f: \X \rightarrow \mbb{R}$ and we assume evaluations of $f(\x)$ are computationally expensive. 
Denote observations of the expensive oracle $f$ as $y_i = y(\x_i),~i=1,2,\ldots, n$. 
Let $X_n$ denote the matrix horizontal stacking of the $n$ observation sites $\x_i,~i=1,\ldots,n$, with the corresponding response vector denoted $\y_n$.  A GP model assumes a multivariate normal distribution over the response, e.g., $\y \sim \mcl{GP}(0, \mbf{K}(X,X))$, where covariance $\mbf{K}(X,X)$ is typically a function of inverse Euclidean distances, i.e., $\mbf{K}(X,X)^{ij} = k(||\x_i - \x_j||^2)$; see \citep{santner2003design,rasmussen:williams:2006,gramacy2020surrogates} for reviews.  Conditioned on observations $\D_n = \{X_n, \y_n\}$, posterior predictions at input $\x$ follow 
\vspace*{0.15cm}
\begin{equation}
    Y(\x) | \D_n \sim \mcl{GP}\left(\mu_n(\x), \sigma^2_n (\x)\right)
    \quad\T{where}\quad
    \begin{aligned}
    \mu_n(\x) &= k(\x, X_n)\mbf{K}(X_n, X_n)^{-1}\y_n \\
    \sigma^2_n(\x) &=  k(\x, X_n) - k(\x, X_n)\mbf{K}(X_n, X_n)^{-1}k(X_n, \x).
    \end{aligned}
    \label{e:GP}
\end{equation}
Throughout, subscript $n$ is used to denote quantities from a surrogate trained on $n$ data points.  The posterior distribution of \Cref{e:GP} provides a general probabilistic surrogate model that can be used to approximate the expensive oracle.

\subsection{Multifidelity Gaussian process regression}
\label{sec:mfgp}
  Crucially, we assume that $f(\x)$ is approximated by models $\hat{f}(\x, s)$, where $\x$ are common inputs to all the models and the parameter $s \in \Sc\subset \mbb{R}$ is a tunable \emph{fidelity} parameter for each model. For the sake of simplicity, in this work, we set $s$ to be scalar-valued and without loss of generality $\Sc= [0,1]$, where $s=1$ and $s=0$ represents the model at the highest and lowest fidelity, respectively. Therefore,
\[ \hat{f}(\x, 1) \defeq f(\x).\]
Additionally, we assume there is a known cost function $c(\x,s): \X\times\Sc \rightarrow \mbb{R}$, which models the computational cost of querying $\hat{f}$ at a specific input-fidelity pair $(\x,s)$. Overall, we are interested in learning $f(\x)$, by querying $\hat{f}$, while keeping the overall computational cost lower than querying $f(\x)$ alone. Our multifidelity method depends on learning a GP model that maps the augmented input-fidelity space $(\x, s) \in \X \times \Sc$ to output quantities of interest $\hat{f}(\x, s)$. 
We assume that we can make noisy evaluations of the function $\hat{f}$ at a given $\x$ and fidelity level $s$. That is
\begin{equation}
    {y}_i = \hat{f}(\x_i, s_i) + \epsilon_i, \quad i= 1,\ldots,n.
\end{equation}
As in the single fidelity setting, we specify GP prior distributions on the
oracle value and the noise, over the joint input-fidelity space. That is, $\hat{f}(\x, s) \sim \mcl{GP}\left(0, k( (\x, s),
(\x', s'))\right)$ and $\epsilon \sim \mcl{GP}(0,\sigma^2_\epsilon)$, where we assume that $\sigma^2_\epsilon$ is a constant but unknown noise variance. 
The covariance function $k$, in this case, captures the correlation between the observations
in the joint $(\x, s)$ space; here we use the product composite form
given by $k((\x, s), (\x', s')) = k_{\x}(\x,\x'; \gamma_\x) \times k_s(s,s';
\gamma_s) $, where $\gamma_\x$ and $\gamma_s$ 
parameterize the covariance functions for $\x$ and $s$, respectively. We estimate the
GP hyperparameters 
$\bs{\Omega}= \lbrace \gamma_\x, \gamma_s, \sigma^2_\epsilon \rbrace$
from data by maximizing the marginal
likelihood~\cite[Ch. 5]{rasmussen:williams:2006}. We choose an anisotropic Matern-type kernel for $k_\x$ which is the defacto choice in GP regression.
The posterior predictive distribution of the output $Y$, conditioned on available observations from the multiple fidelity oracles, is given by~\cite{rasmussen:williams:2006} 
\begin{equation}
    \begin{split}
        Y(\x, s) |
      \D_n, \bm{\Omega} &\sim \mcl{GP}(\mu_n(\x, s), \sigma^2_n (\x, s)),~ \quad \D_n \defeq  \lbrace
(\x_i, s_i), {y}_i \rbrace _{i=1}^n \\
\mu_n(\x, s) &= k((\x, s), X_n)^\top [\mbf{K}_n + \sigma_\epsilon^2 \mbf{I}]^{-1} \mbf{y}_n\\
\sigma^2_n(\x, s) &=  k((\x, s), (\x, s)) - k((\x, s), X_n)^\top [\mbf{K}_n + \sigma_\epsilon^2 \mbf{I}]^{-1} k((\x, s), X_n),
    \end{split}
    \label{e:GP}
\end{equation}
where with a slight abuse of notation, we use $X_n$ to denote the row-stack of input fidelity pairs 
$[\x_i^\top, s_i],~\forall i=1,\ldots,n$, and $\mbf{y}_n$ is the vector of all
observations at various fidelty levels in $\D_n$. Note that other notations remain identical to the single-fidelity GP. For instance, $\mu_n$ and $\sigma_n^2$
are the posterior mean and variance of the multifidelity GP, respectively, where the
subscript $n$ implies the conditioning based on $n$ past observations. Likewise, the training of the multifidelity GP is also via marginal likelihood maximization.
\subsection{Active learning}
The GP posterior distribution serves as a multifidelity surrogate model for the expensive quantity of interest derived from the high-fidelity model. The accuracy of the surrogate model is naturally dependent on the observed data used in the model fitting and it is impossible to choose the observations apriori such that they would result in a sufficiently accurate surrogate model. Therefore, we are interested in first constructing the surrogate model with a (randomly chosen) set of \emph{seed} observation samples, which is then adaptively improved by adding new observations that are judiciously chosen according to a certain criterion. This adaptive model improvement can be achieved by defining an \emph{acquisition function}, $\alpha(\x,s)$, in terms of the GP posterior, that quantifies the utility of a candidate input-fidelity pair $(\x,s)$. This way, new queries to the simulation model can be made at optimal $(\x,s)$ values that maximize $\alpha(\x,s)$, that is,
\begin{equation}
    \x_{n+1}, s_{n+1} = \argmax_{(\x,s) \in \X\times \Sc}~\alpha(\x, s) | \D_n
\end{equation}
It should be noted that the acquisition function optimization is relatively computationally cheap, since it depends on the GP posterior and is independent of the high-fidelity model. A generic active learning algorithm, in the single fidelity setting, is shown in \Cref{a:BO}.

There are several choices one could make for $\alpha$, when the overall goal is model predictive accuracy. As previously mentioned, we explore leave-one-out cross-validation \cite{wahba1985comparison}, \citep[Ch.5]{rasmussen:williams:2006}. But, there exist other approaches which include expected improvement for global fit \cite{lam2008sequential} and the integrated mean-squared predictive error (or posterior variance) reduction by \cite{seo2000gaussian,chen2014sequential,binois2019replication}. To the best of our knowledge, none of the existing work automatically extend to a multifidelity setting and neither does a multifidelity active learning method exists in the context of LOO-CV. We now proceed to introduce the proposed \mfcv~method, that is a novel active learning approach with multifidelity GPs and LOO-CV.

\begin{algorithm2e}[t]
\textbf{Given:} $\D_n = \lbrace
(\x_i, s_i), {y}_i \rbrace _{i=1}^n$, 
 total budget $B$, 
 and GP hyperparameters $\bm{\Omega}$ \\
\KwResult{Posterior GP $\mcl{GP}(\mu_B(\x), \sigma_B^2(\x))$}
  \For{$i=n+1, \ldots, B$, }{
  Find $\x_i \in \underset{\x \in \X}{\argmax}~ \alpha(\x)$ \qquad (acquisition function maximization)\\
    Observe ${y}_i$ = $f(\x_i) + \epsilon_i$\\ 
    Append $\D_i = \D_{i-1} \cup \lbrace (\x_i), {y}_i \rbrace$\\ 
    Update GP hyperparameters $\bm{\Omega}$ \\
 }
 \caption{Generic Gaussian process active learning}
 \label{a:BO}
\end{algorithm2e}

\section{Proposed method}
\label{sec:proposed_method}

\subsection{Multifidelity cross-validation (\mfcv)}
\label{subsec:mfcv}
Because of the marginalization properties of GPs, $Y(\x,1) | \D_n$ is also a GP and can be expressed as $Y(\x,1) \sim \mcl{GP}(\mu_n(\x,1), \sigma_n^2(\x,1))$. Now we use the concept of $k$-fold cross-validation, where the observed data is split into $k$ disjoint, equally sized subsets. Then the GP is trained on $k-1$ subsets while the validation is done on the remaining subset. This procedure is repeated $k$ times to rotate amongst the subsets. An extreme case of $k$-fold cross-validation is when $k=n$, that is particularly suited for small-data situations and is called the leave-one-out cross-validation (LOO-CV). Fortunately, the GP posterior and the marginal log likelihood, with one data point left out from $\D_n$, is available in closed-form enabling efficient computation~\cite{rasmussen:williams:2006}; we present that in what follows. The posterior predictive distribution at $y_i$, when $(\x_i, s_i)$ are left out, is given by
\begin{equation}
    p(y_i|(\x, s), \mbf{y}_{-i}, \bs{\Omega} ) \sim \mcl{N} (\mu_{-i}(\x, s), \sigma_{-i}^2(\x, s)),
\end{equation}
where $\mbf{y}_{-i}$ refers to all observations except the $i$th and likewise, $\mu_{-i}$ and $\sigma^2_{-i}$ are the GP posterior mean and variance conditioned on $\mbf{y}_{-i}$. The LOO-CV predictive posterior mean and variance are given by the following closed-form expressions~\cite{rasmussen:williams:2006}
\begin{equation}
    \begin{split}
        \mu_{-i} =& \hat{y}_i - \f{[\mbf{K}^{-1} \mbf{y}]_i}{[\mbf{K}^{-1}]_{ii}} \\
        \sigma^2_{-i} =& \f{1}{[\mbf{K}^{-1}]_{ii}}
    \end{split}
    \label{eqn:loocv_mv}
\end{equation}
where $[]_i$ and $[]_{ii}$ denote the $i$th element of a vector and the $(i,i)$th element of a matrix, respectively. Note that once $\mbf{K}^{-1}$ is computed, computing \eqref{eqn:loocv_mv} takes only $\mcl{O}(n^2)$ operations. In fact, in \Cref{eqn:loocv_mv}, the $\mbf{K}^{-1}$ is never explicitly computed. Once $\mbf{K}$ is computed, we compute $\mbf{K}^{-1} \y$ and $[\mbf{K}]_{ii}$ using the method of  linear conjugate gradients~\citep[Ch. 5]{nocedal1999numerical}.

\newcommand{\ecv}{e_\T{cv}}
Let $\ecv(\x,s)$ denote the leave-one-out cross-validation error everywhere in $\X\times\Sc$. Then,
the main idea of the proposed approach is to learn $\ecv(\x,s)$ so that our original posterior GP, $Y(\x,s)$,
can be adaptively enriched by adding points that drive $\ecv(\x,1)$ to zero everywhere in $\X\times \{1\}$. In other words, we will choose points that maximize $\ecv(\x',1)$, due to a candidate input-fidelity pair $(\x,s)$, at every step of the adaptive model building process. 
Learning $\ecv$ begins by making the following (stochastic) observations
\begin{equation}
    e_{\T{cv},i} = (Y_{-i}(\x_i, s_i) - {y}_i)^2,~\forall i=1, \ldots, n,
\end{equation}
where $Y_{-i}$ is the posterior GP conditioned on $\D_{-i}$ at $(\x_i, s_i)$ and is a normal random variable with mean $\mu_{-i}$ and variance $\sigma^2_{-i}$. $\ecv$, on the other hand, assumes a noncentral chi-squared distribution with one ($\kappa=1$) degrees of freedom, with the following mean and variance:
\begin{equation}
    \begin{split}
        e_{\T{cv}, i} \sim & \chi^2 \left( \kappa = 1, \lambda = (\mu_{-i}(\x_i, s_i) - y_i)^2 \right) \\
        \mbb{E}[e_{\T{cv}, i}] =& 1 + (\mu_{-i}(\x_i, s_i) - y_i)^2 \\
        \mbb{V}[e_{\T{cv}, i}] =& 2(1 + 2(\mu_{-i}(\x_i, s_i) - y_i)^2),
    \end{split}
    \label{eqn:chisq}
\end{equation}
where $\lambda$ is the noncentrality parameter. The expectation and variance in \Cref{eqn:chisq} are clearly dependent only on the posterior means computed in \Cref{eqn:loocv_mv} and hence can be readily computed. In principle, the mean and variance in \eqref{eqn:chisq} can be used to train a heteroscedastic GP. However, in this work, as a first step, we take only the expected value of $\ecv$, at every $i$, and fit a GP to it. In order to comply with the infinite support of GPs, we instead fit the GP to the logarithm of the expected value of $\ecv$.
We define the logarithm of the $e_\T{cv}$ and place a GP prior on it as follows
\begin{equation}
    \ell(\x,s) = \log(e_{cv}(\x,s)) \sim \mcl{GP}(0, k_\ell(\cdot,\cdot)),
\end{equation}
whis learned from observations  $\log (\mbb{E}[e_{\T{cv}, i}]),~\forall i = 1,\ldots,n$. As before, the posterior predictive distribution of $\ell(\x,s)$ conditioned on the observed LOO-CV error is also a GP.  Finally, the \mfcv~acquisition function is given as 
\begin{equation}
    \alpha_\T{MFCV}(\x,s) = \mbb{E}_\ell \left[ \max_{\x' \in \X}  \ell(\x', 1)|\D_n \bigcup \{(\x,s), \ell(\x, s)\} \right].
    \label{eqn:mfcv}
\end{equation}
Our proposed \mfcv~acquisition function, shown in \eqref{eqn:mfcv}, quantifies the utility of a given input-fidelity pair $(\x,s)$, on the maximum cross-validation error at the highest fidelity ($s=1$). Then, the new point is chosen as
\begin{equation}
    \x_{n+1}, s_{n+1} = \argmax_{(\x,s) \in \X\times \Sc}~\f{\alpha_\T{MFCV}(\x, s)}{c(\x,s)}.
\end{equation}

Our approach equally applies to selecting a batch of $q$ points. Let $X = \{\x^1, \ldots, \x^q\}$, $S = \{s^1, \ldots, s^q\}$, and $\boldsymbol{\ell} = \{\ell^1, \ldots, \ell^q\}$, then our acquisition function in batch mode is given by
\begin{equation}
    \alpha_{q\T{MFCV}}(X, S) = \mbb{E}_{\boldsymbol{\ell}} \left[ \max_{\x' \in \X}  \ell(\x', 1)|\D_n \bigcup \{(X, S), \boldsymbol{\ell} \} \right],
    \label{eqn:qmfcv}
\end{equation}
and as before, the next batch of points is chosen as
\begin{equation}
   X_{n+1:q}, S_{n+1:q} = \argmax_{(X, S) \in \X^q\times \Sc^q}~\f{\alpha_{q\T{MFCV}}(X,S)}{\sum_{i=1}^{q}c(\x^i,s^i)}.
\end{equation}
Plugging in \eqref{eqn:mfcv} and \eqref{eqn:qmfcv} into \Cref{a:BO} (line 3) completes our method. The overall \mfcv~algorithm is summarized in \Cref{a:MFCV}.

\begin{algorithm2e}[t]
\textbf{Given:} Seed observations $\D_n = \lbrace
(\x_i, s_i), ({y}_i,\ldots,{y}_n) \rbrace _{i=1}^n$, 
 total number of iterations $B$, cost model $c(\cdot, \cdot)$
 and GP hyperparameters $\bs{\Omega}$ \\
\KwResult{Posterior GP at $s=1$, $\mcl{GP}(\mu_B(\x, 1), \sigma_B^2(\x, 1))$}
  \For{$i=n+1, \ldots, B$, }{
  Compute $\ecv$ at all observation sites: $\mbb{E}[e_{\T{CV}, k}],~k=1,\ldots,i$. \\
  Fit an inner GP to the data set $\{(\x_k, s_k), \log(\mbb{E}[e_{\T{CV}, k})\},~k=1,\ldots,i$.\\ 
  Compute and optimize the cost-aware \mfcv~acquisition function (\Cref{eqn:mfcv}). \\
  Make new observations $y_i = \hat{f}(\x_i, s_i)$. \\
  Append $\D_{i} = \D_{i} \cup \lbrace (\x_i, s_i), {y}_{i} \rbrace$\\ 
  Update GP hyperparameters $\bs{\Omega}$ \\
 }
 \caption{Multifidelity cross-validation (\mfcv)}
 \label{a:MFCV}
\end{algorithm2e}

\section{Numerical Experiments}
\label{sec:num_expts}

We first demonstrate our proposed methods on synthetic functions of varying dimensions as well as one real-world experiment. For all the experiments, we provide a set of seed points $\mcl{D}_n = \{(\x_i, s_i), y_i,~ i=1,\ldots,n\}$ to start the algorithm, where $n=10d$ and $(\x_i, s_i), \forall i$ are selected uniformly at random from $\X \times \Sc$. We use the root mean-squared error (RMSE) as the main metric to assess the method, which is defined as
\[\T{RMSE} = \sqrt{\frac{ \left(\mu_n(\x, 1) - f(\x) \right)^2}{N}}.\]
The RMSE is evaluated on a set of test points chosen uniformly at random, with cardinality $N$, chosen at the highest fidelity $s=1$; $N$ is set to be $30d$. We keep the test point fixed across iterations to ensure the error is always evaluated against a constant benchmark. We observe the history of RMSE against the cumulative cost of evaluations.
Since, at the time of writing, we are not aware of any other competing method for multifidelity cross-validation, we primarily compare our method against the single-fidelity equivalent of our method---that is, acquisitions performed at $s=1$; we denote this as HF. We look at three variants of MFCV: sequential sampling $(q=1)$, and batch sampling $q=2,~4$. Finally, we compare our method against a "random" selection strategy according to Sobol sampling~\cite{sobol1967distribution}---this way, we ensure there is value in choosing acquisitions that are optimal according to our acquisition function. All the experiments, including the HF, are provided with exactly the same set of seed points. Once the algorithm begins, all experiments choose an input-fidelity pair $(\x, s)$, whereas the HF will do so at a fixed fidelity of $s=1$.
All experiments are repeated 10 times, but for the turbine experiment which is repeated 5 times, with random seed points, and we look at the average performance metric along with $\pm 1$ standard deviations. 

\begin{figure}
    \centering
    \includegraphics[width=.7\textwidth]{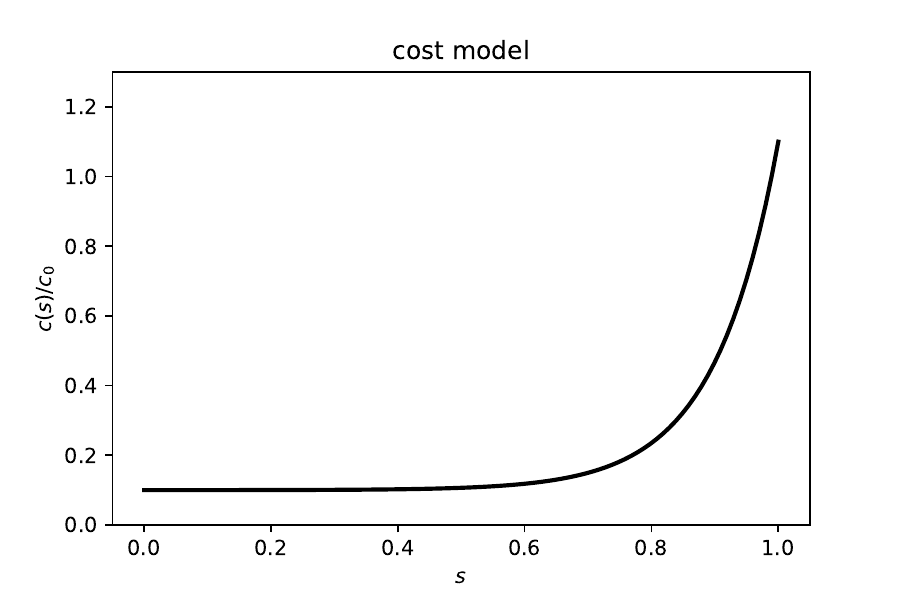}
    \caption{Multifidelity cost model $c(s)$ with $c_1 = 10$, which emphasizes the cost disparity across fidelities as one approaches $s=1$.}
    \label{fig:cost_model}
\end{figure}

For all the experiments, the following cost model is assumed:
\[ c(s) / c_0 = c_2 + \exp\left(-c_1 \times (1 - s) \right),\]
where $c_0$ is a normalizing constant that is required to ensure that the acquisition function optimization is not biased by the actual costs of the models; note that $c_0$ is also the cost of the high-fidelity ($s=1$) model, for which we fix $c_0 = 500$. The $c_1>0$ determines the steepness of the cost model; we set $c_1 = 10$ for all the experiments. Finally, $c_2> 0$ is a constant to ensure that the cost model is bounded away from zero and is roughly the cost of the lowest-fidelity model ($\hat{f}(\cdot, 0)$); we fix $c_2 = 0.1$. We note that our cost model---by virtue of setting $c_1=10$---is designed to emphasize the disparity between the computational costs of lower- and higher-fidelity model queries; see \Cref{fig:cost_model}. Note that we assume the cost model is independent of $\x$ for simplicity and without any loss of generality.

\subsection{Synthetic experiments}
\label{s:expts_synth}
\begin{figure}[htb!]
    \centering
    \begin{subfigure}{.5\textwidth}
        \includegraphics[width=1\linewidth]{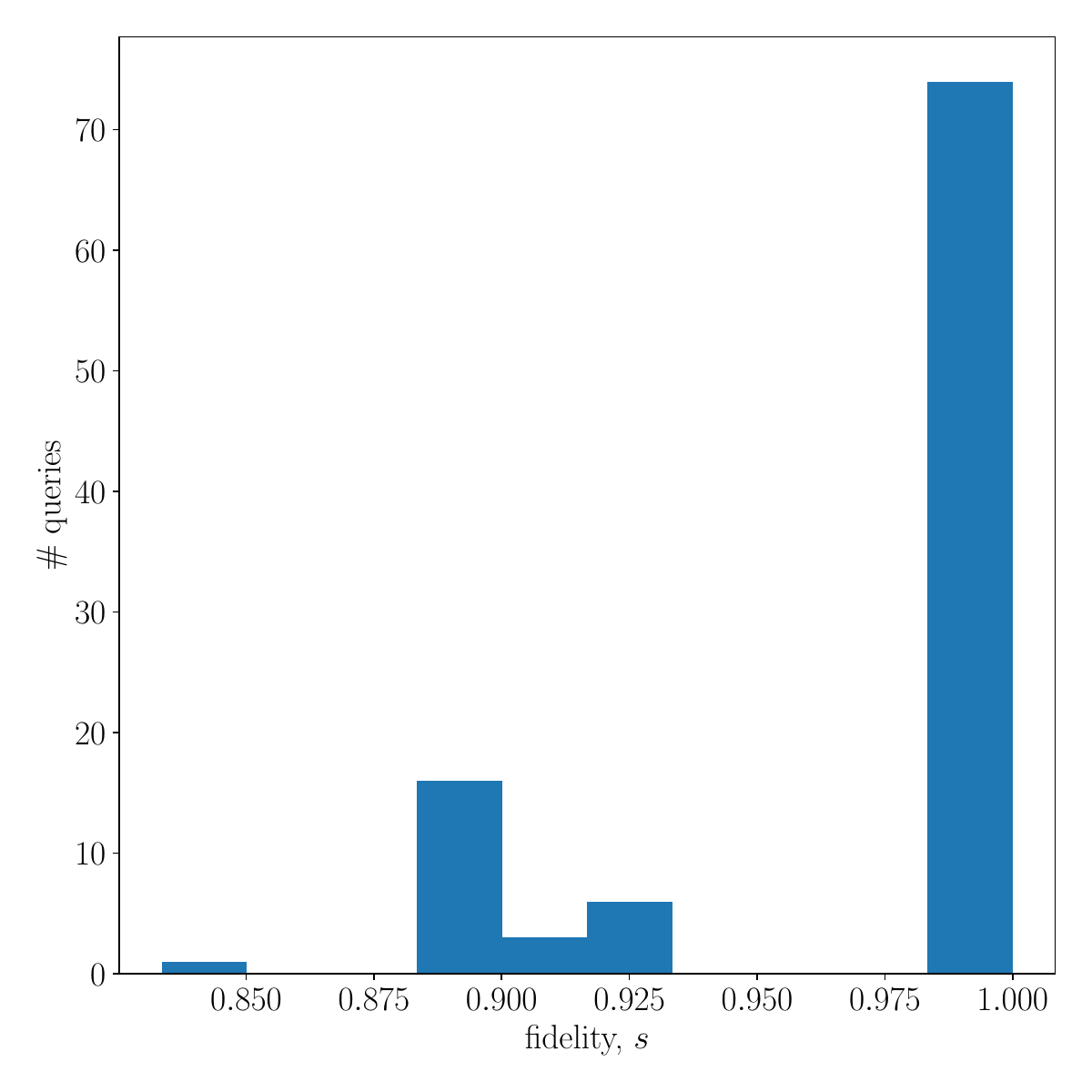}
        \caption{Branin (3D)}
    \end{subfigure}%
    \begin{subfigure}{.5\textwidth}
        \includegraphics[width=1\linewidth]{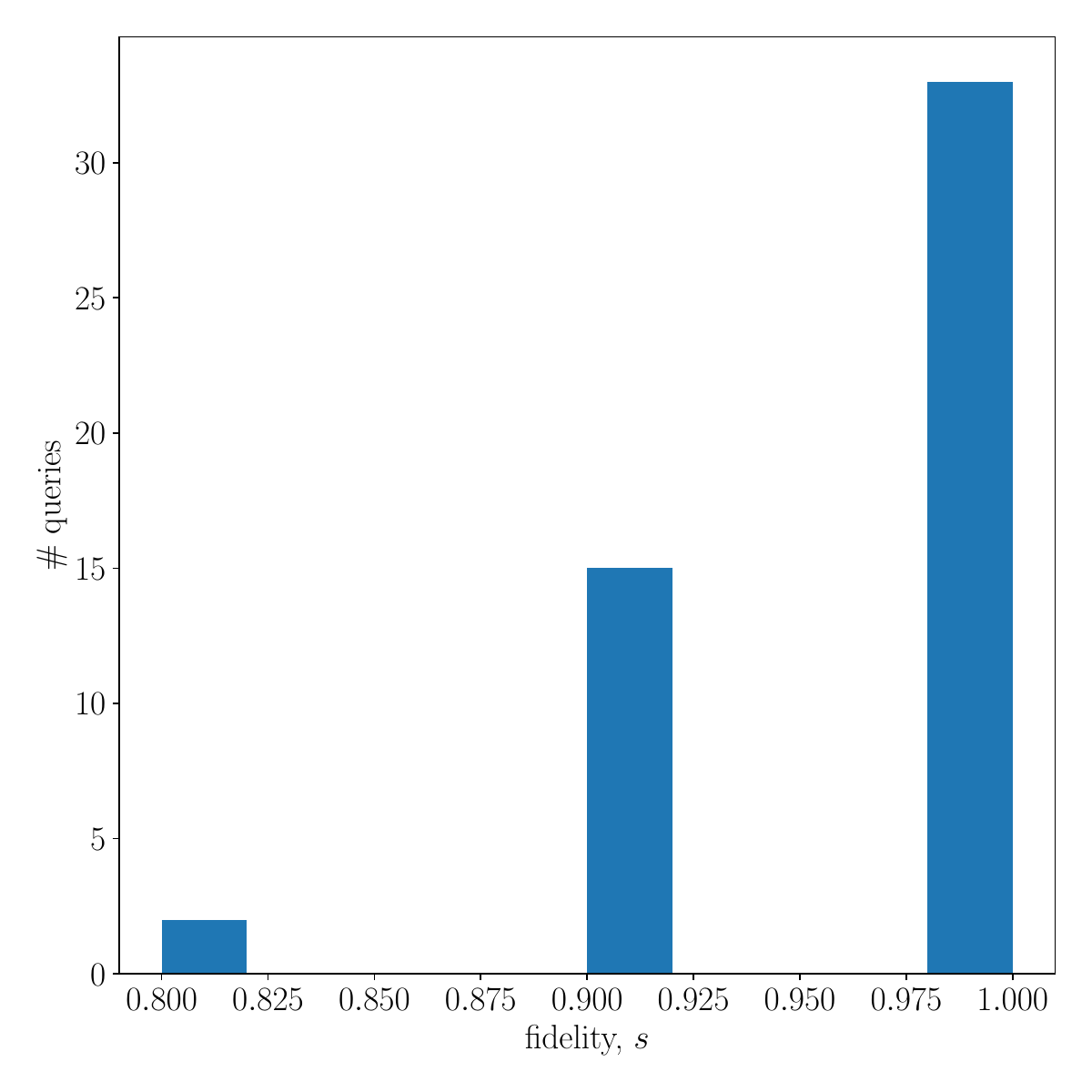}
        \caption{Four branches (3D)}
    \end{subfigure} \\
    \begin{subfigure}{.5\textwidth}
        \includegraphics[width=1\linewidth]{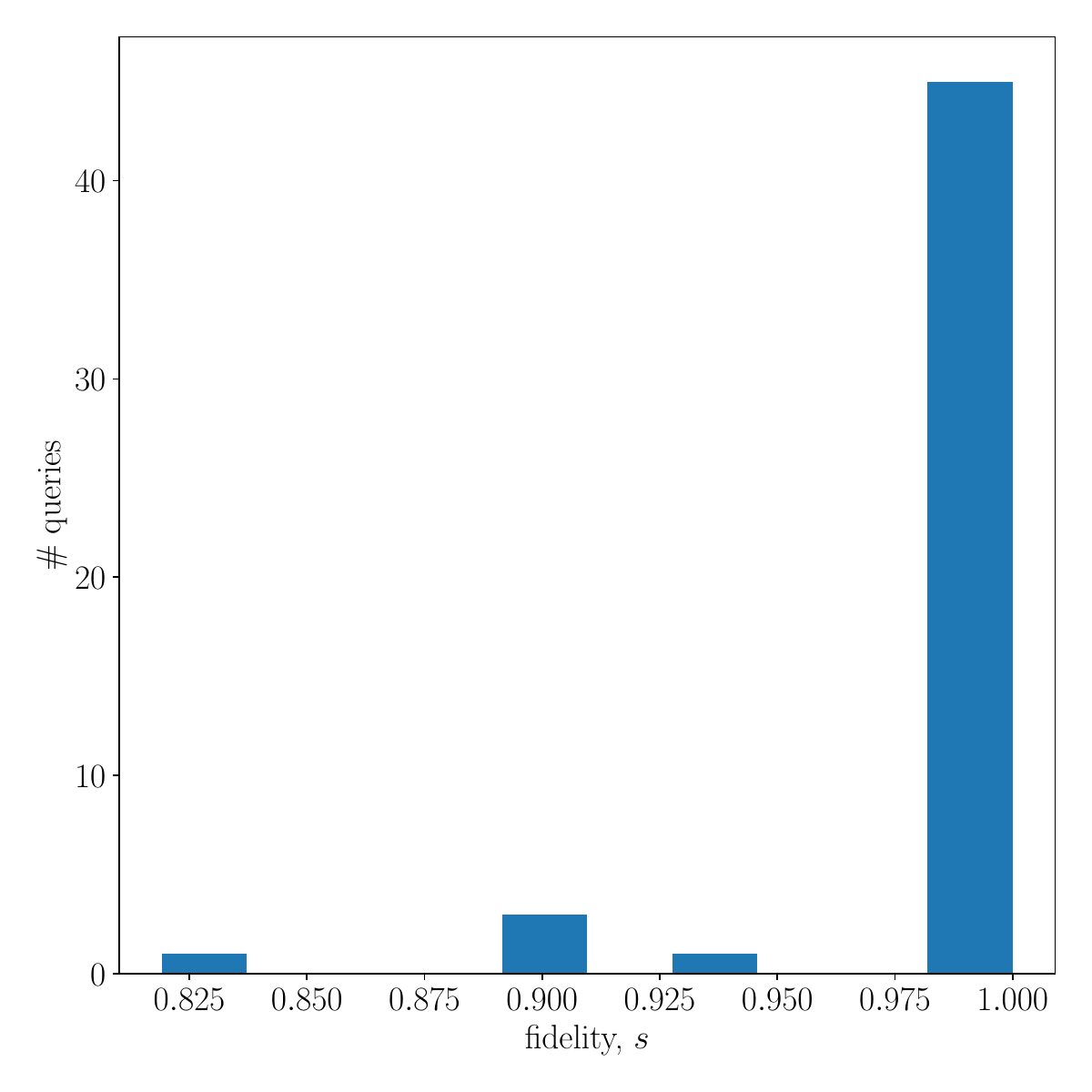}
        \caption{Ishigami (4D)}
    \end{subfigure}%
    \begin{subfigure}{.5\textwidth}
        \includegraphics[width=1\linewidth]{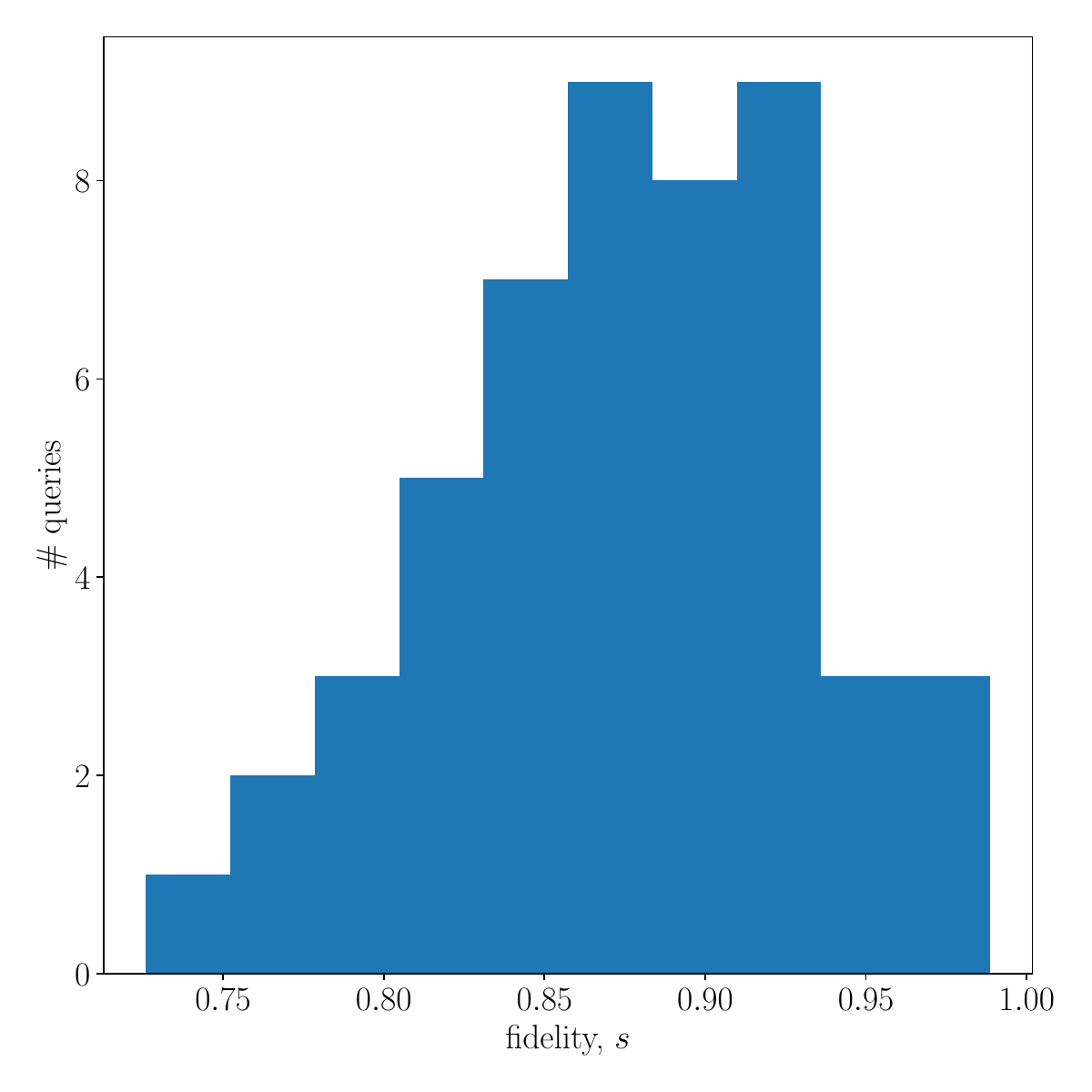}
        \caption{Hartmann (7D)}
    \end{subfigure} 
    \caption{Fidelity levels selected by the MFCV strategy. Note that we show only $q=1$.}
    \label{fig:synthetic_fidelities}
\end{figure}

We demonstrate our method on four synthetic test functions: multifidelity Branin ($d=3)$, multifidelity four branches $d=3$, multifidelity Ishigami~\cite{renganathan2023camera} ($d=4$), and multifidelity Hartmann ($d=7$).
The results for the synthetic test functions are shown in \Cref{fig:synthetic}; details and plots of these functions are shown in the appendix. The plot shows the average RMSE of all repetitions (solid line) along with the $\pm 1$ standard deviation (shaded), versus the cumulative computational cost. Note that we show the RMSE starting with the first iteration, that is not including the seed points. Therefore, since we show the plot versus the cumulative cost, the lines within each plot do not start from the same point, despite starting with exactly the same seed points. Notice that some variant of the \mfcv ($q=1,2, \T{or},~4$) always outperforms the single-fidelity HF case, thereby clearly demonstrating the benefit of the multifidelity approach. Furthermore, the random strategy (Sobol) is consistently worse than all other strategies. This demonstrates that the proposed acquisition strategy has value of information and the points chosen are indeed optimal according to the acquisition function. It should be noted that the plots of RMSE don't show a monotonic decrease in RMSE. This is because, with every new acquisition, the hyperparameters of the global GP are computed again, which involves a highly nonconvex optimization of the marginal log likelihood function with randomized starting points. As a result, the posterior GP is not guaranteed to result in a GP with improved RMSE, although the overall trend of these curves show a decreasing trend as expected. This phenomenon is consistent with other results reported in the literature~\cite{mohammadi2022cross}. In \Cref{fig:synthetic_fidelities}, we show the distribution of the fidelity levels selected by \mfcv. The distribution, as evident from the figure, is unique to each experiment. However, in this case, fidelities $\geq 0.7$ are only selected, showing that lower fidelities contribute relatively less in reducing the $\ecv$ at the highest fidelity $s=1$. This is particularly true because the $\ecv$ is heavily influenced by the GP posterior variance and our chosen, stationary, GP prior correlates less when the Euclidean distance between points is high. 

\begin{figure}[htb!]
    \centering
    \begin{subfigure}{.5\textwidth}
        \includegraphics[width=1\linewidth]{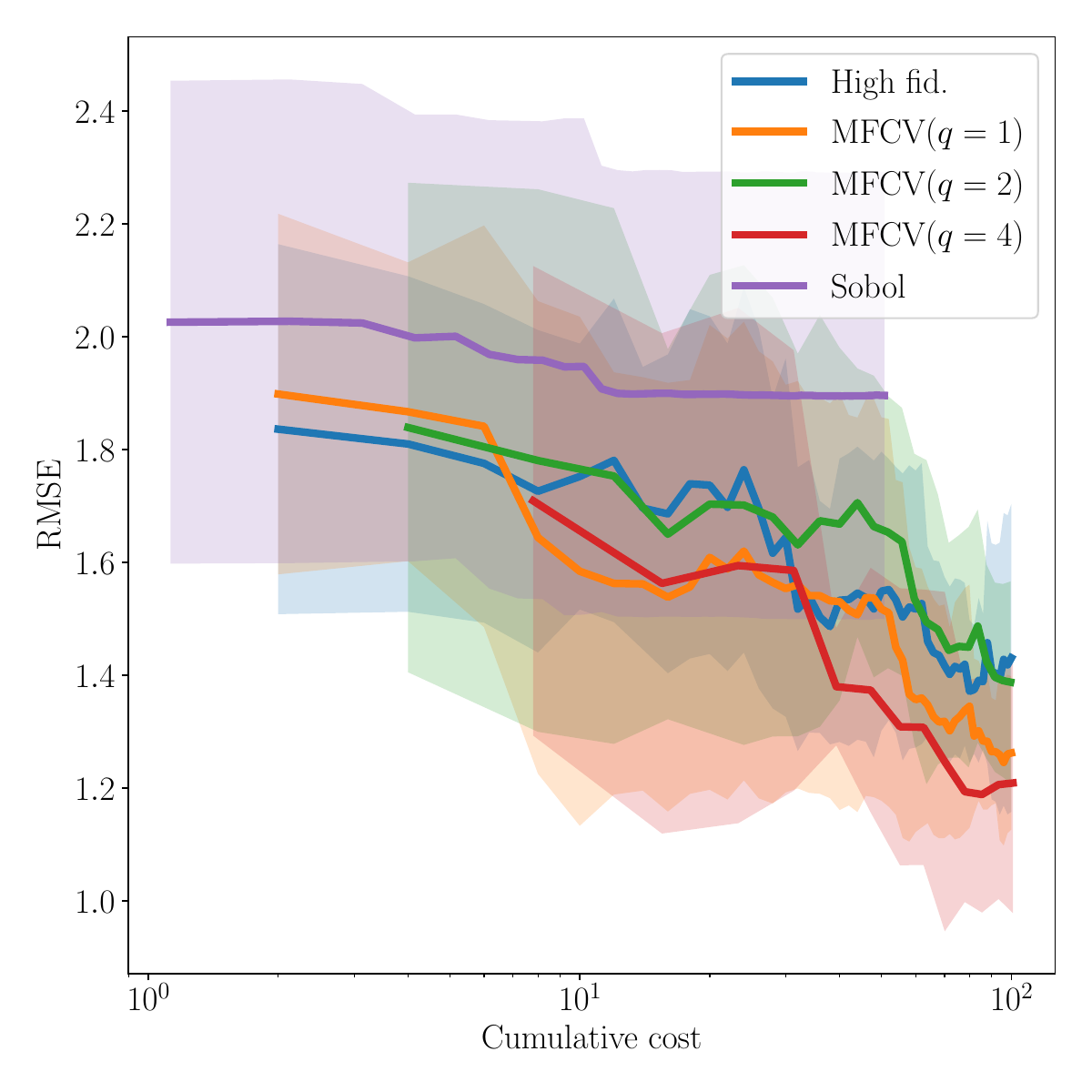}
        \caption{Branin (3D)}
    \end{subfigure}%
    \begin{subfigure}{.5\textwidth}
        \includegraphics[width=1\linewidth]{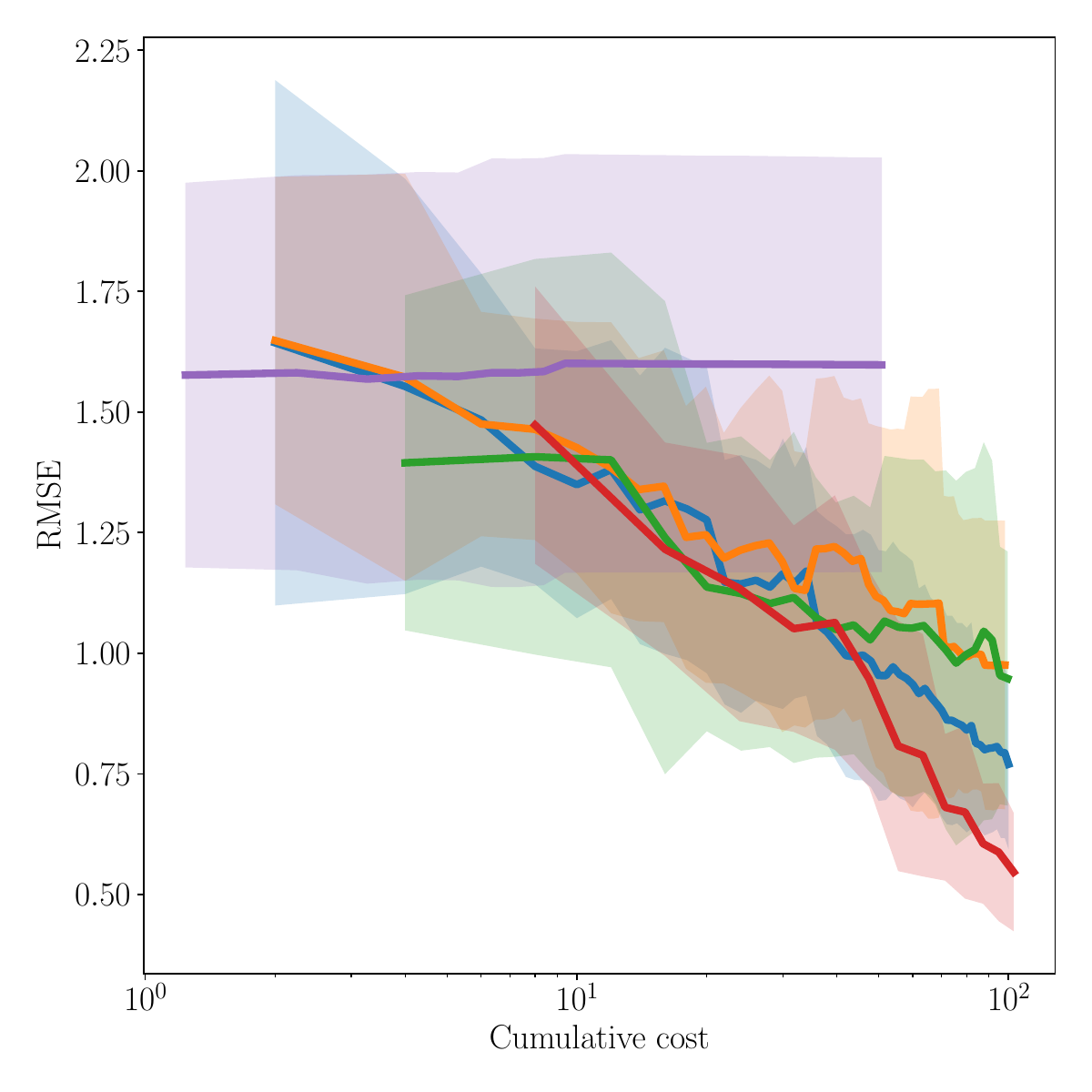}
        \caption{Four branches (3D)}
    \end{subfigure} \\
    \begin{subfigure}{.5\textwidth}
        \includegraphics[width=1\linewidth]{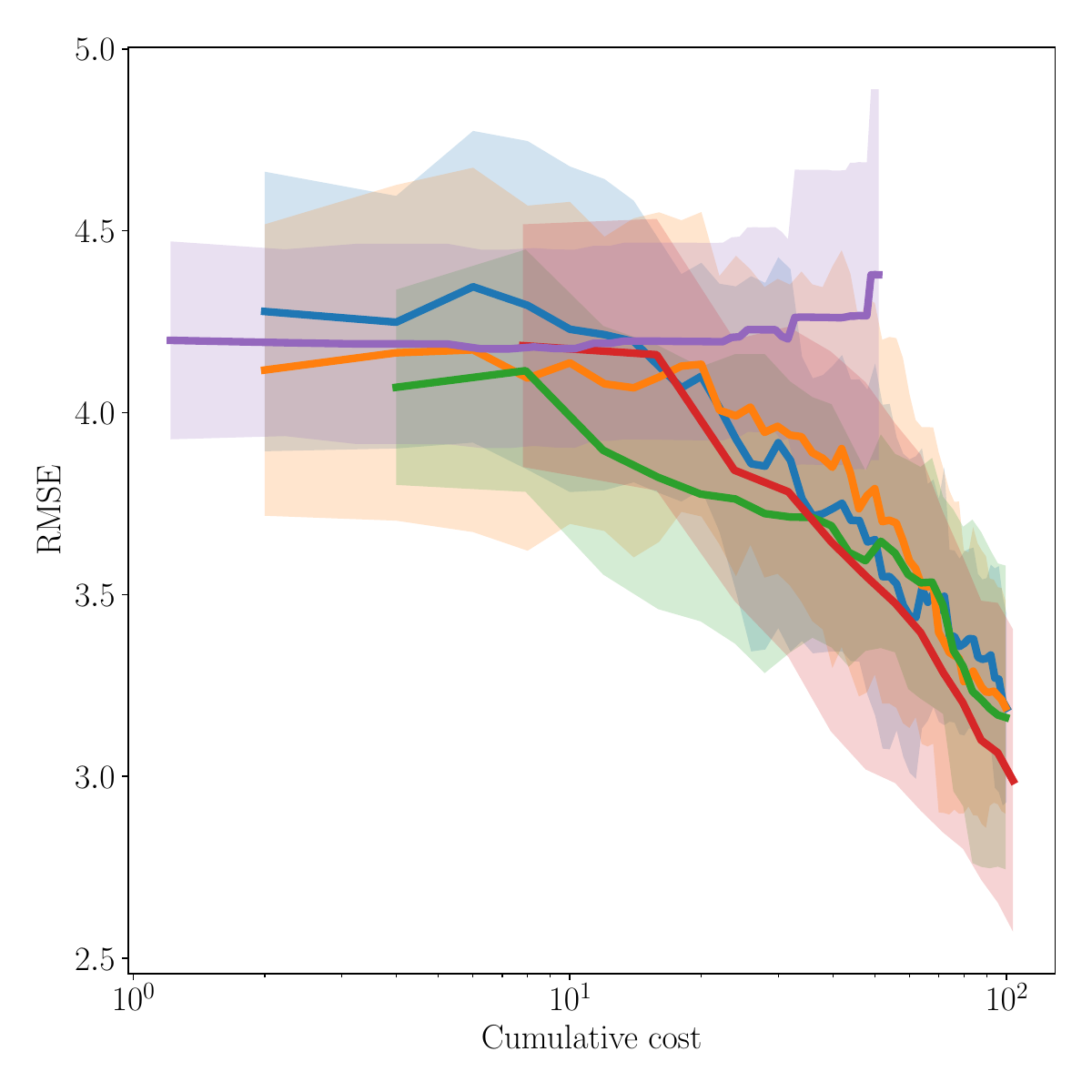}
        \caption{Ishigami (4D)}
    \end{subfigure}%
    \begin{subfigure}{.5\textwidth}
        \includegraphics[width=1\linewidth]{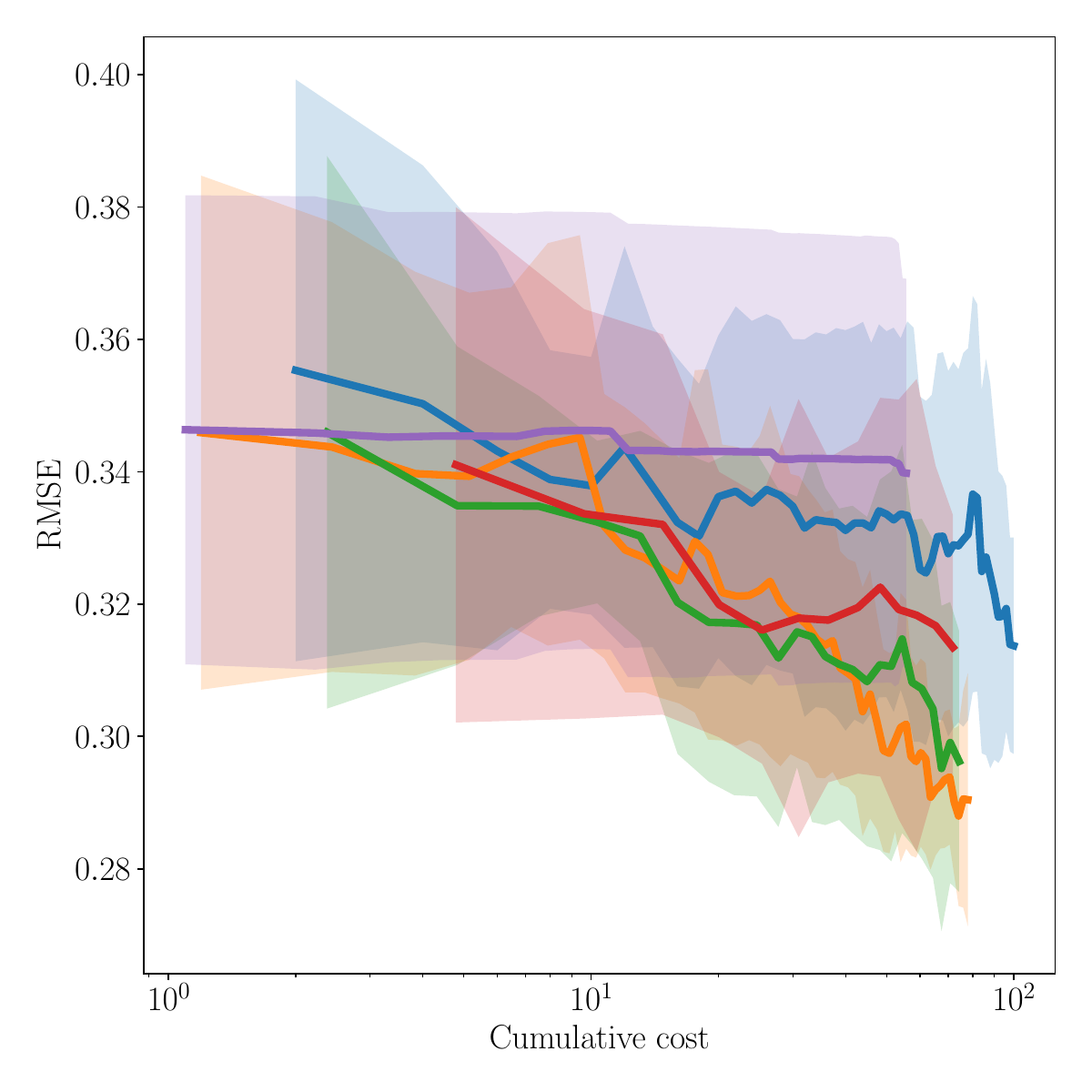}
        \caption{Hartmann (7D)}
    \end{subfigure} 
    \caption{\mfcv~demonstration on synthetic test functions.}
    \label{fig:synthetic}
\end{figure}

\subsection{Gas turbine blade thermal stress analysis on a discrete fidelity space}
\label{sec:turbine}
We now demonstrate our method on the structural analysis of a gas turbine blade at steady-state operating conditions. Gas turbine blades are subject to very high pressures and temperatures that can cause concentrated zones of high stress that can in turn lead to structural deformation and potential catastrophic damage of the engine parts. A finite element model of the blade is leveraged to simulate the thermal stress acting on the bade surface. The Von Mises stress is a good indicator for the distribution of the thermal stress on the blade surface. The maximum Von Mises stress on the blade surface could be used to predict the likelihood of a structural failure~\cite{renganathan2023camera}.

\begin{figure}[h!]
    \centering
    \begin{subfigure}{.34\textwidth}
        \centering
        \includegraphics[width=1\linewidth]{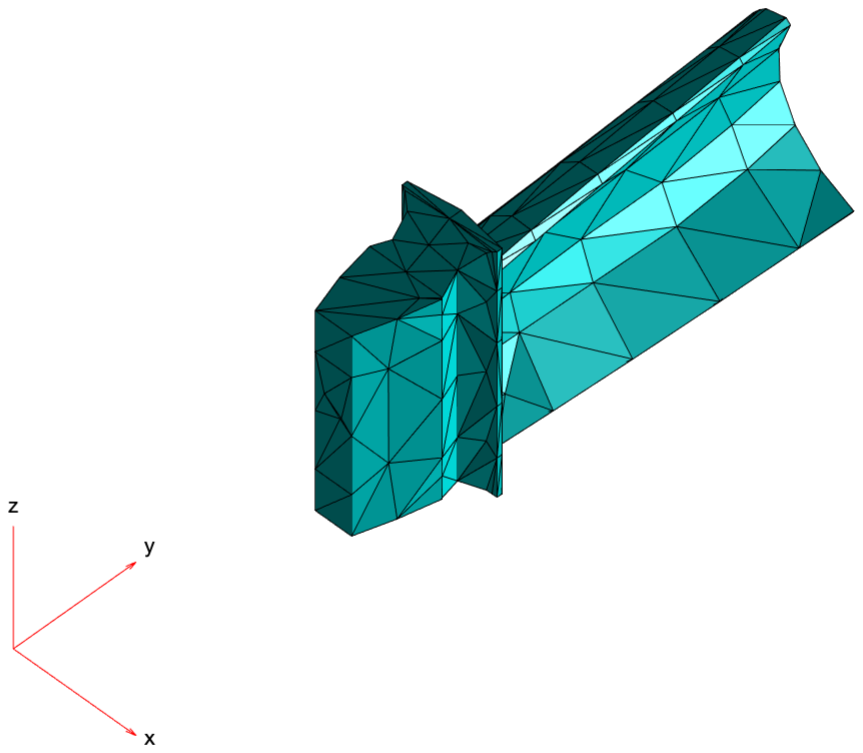}
        \caption{Low fidelity ($s=0.0$)}
    \end{subfigure}%
    \begin{subfigure}{.34\textwidth}
    \centering
        \includegraphics[width=1\linewidth]{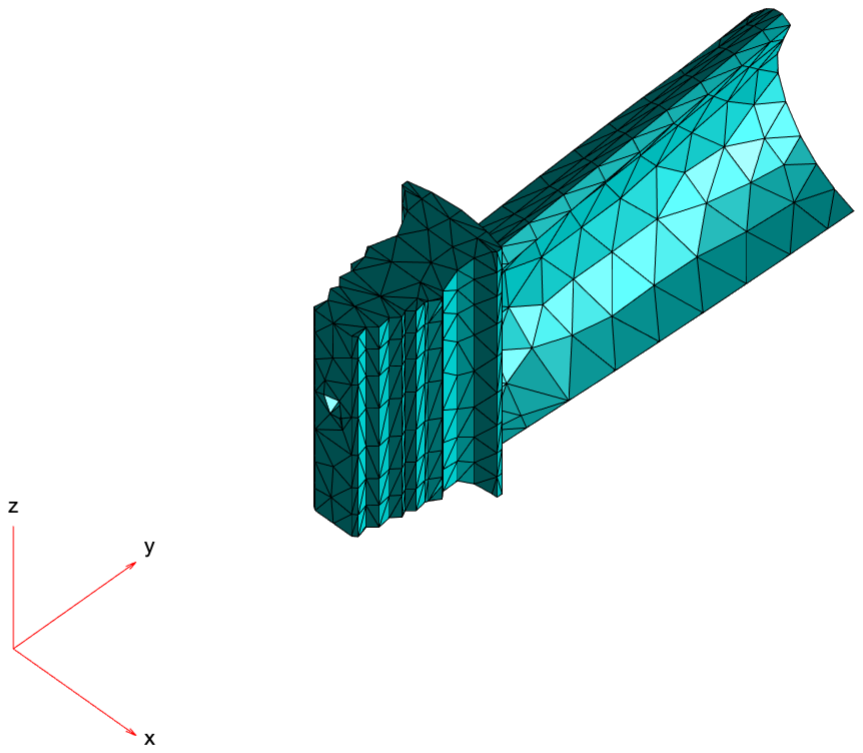}
        \caption{Medium fidelity ($s=0.5$)}
    \end{subfigure}%
    \begin{subfigure}{.34\textwidth}
        \includegraphics[width=1\linewidth]{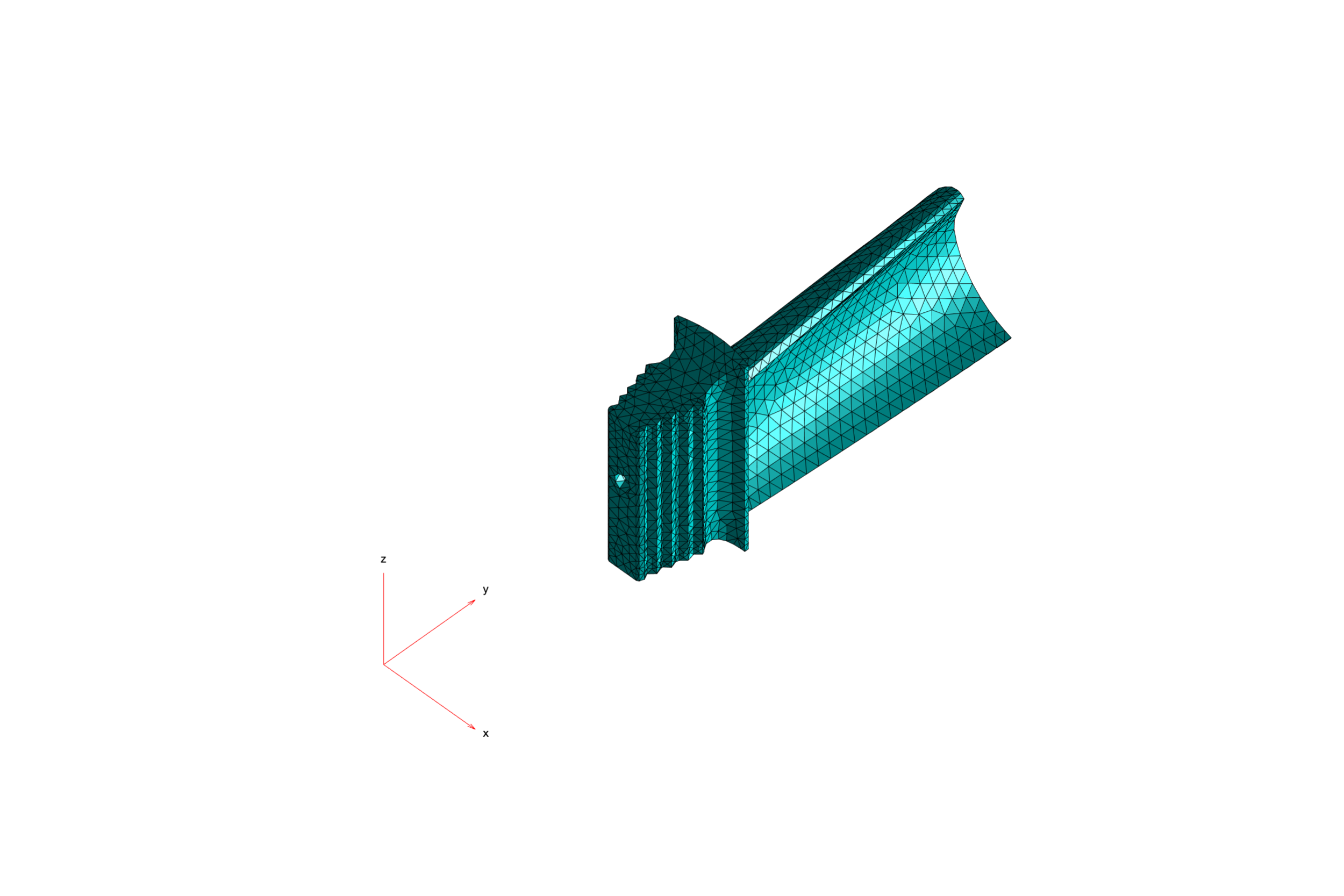}
        \caption{High fidelity ($s=1.0$)}
    \end{subfigure}%
    \caption{Mesh resolutions for the three discrete fidelities levels.}
    \label{fig:turbine_fidelities}
\end{figure}

\begin{figure}
    \centering
    \begin{subfigure}{.5\textwidth}
        \centering
        \includegraphics[width=1\linewidth]{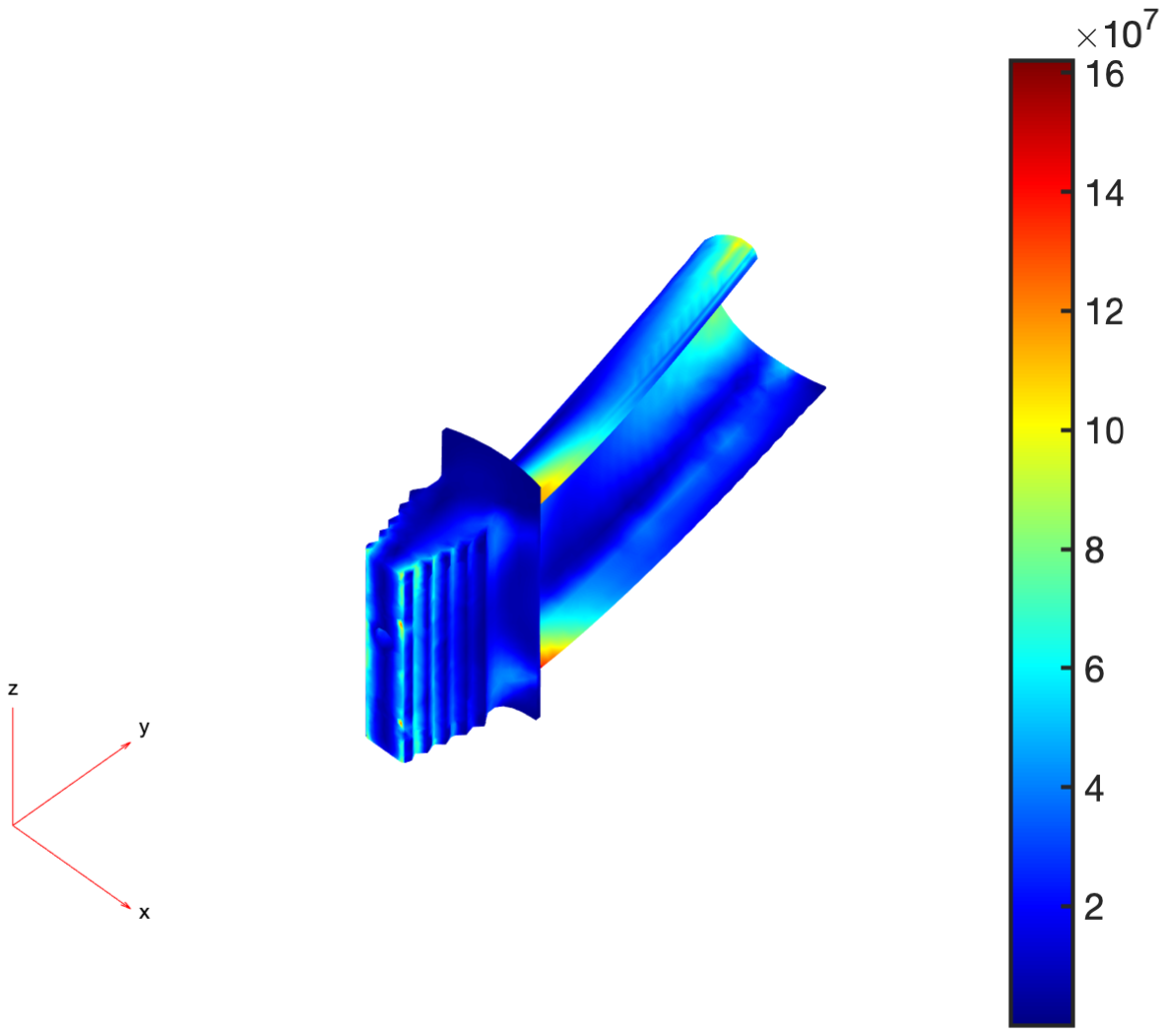}
        \caption{Solution at $\x = [7E5, 4.5E5]$}
    \end{subfigure}%
    \begin{subfigure}{.5\textwidth}
    \centering
        \includegraphics[width=1\linewidth]{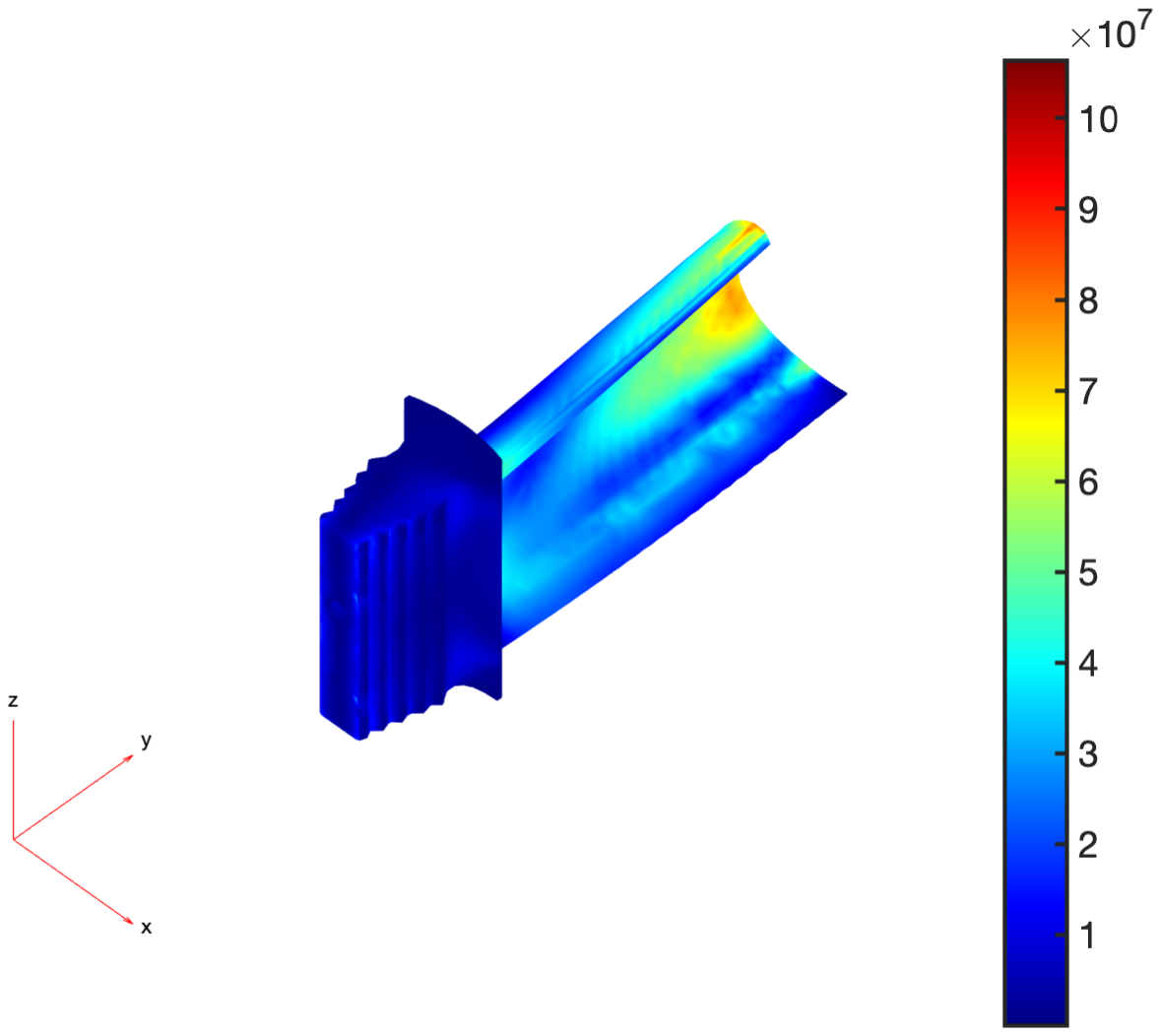}
        \caption{Solution at $\x = [5E5, 4.5E5]$}
    \end{subfigure}%
    \caption{Von Mises stress distribution and deformation on the gas turbine blade, under two different pressure loading boundary conditions.}
    \label{fig:turbine_solution}
\end{figure}

In this regard we vary the pressure-side ($x_1$) and suction-side ($x_2$) pressures as boundary conditions, and set the surface mesh resolution to capture the fidelity; see \Cref{fig:turbine_fidelities} for the three levels of fidelity. Crucially, this experiment demonstrates the methodology on a discrete fidelity set. We observe the maximum Von Mises stress acting on the blade as our response.  The blade material properties are fixed to the following constants: Young's modulus $227E+9$ Pa, Poisson's ratio $0.27$, and coefficient of thermal expansion  $12.7E-6$. The governing equations are solved using a finite element method using Matlab's PDE toolbox~\footnote{https://www.mathworks.com/products/pde.html}. The computational mesh density is determined via a minimum element size which also serves as the fidelity parameter. Specifically, we use the minimum mesh element sizes $\{0.05, 0.025, 0.01\}$, which correspond to fidelities $\mcl{S} = \{0.0, 0.5, 1.0\}$, respectively. The mesh resolutions for the specified three fidelity levels are shown in \Cref{fig:turbine_fidelities}. As previously mentioned, the maximum von Mises stress is the quantity of interest in this experiment; see \Cref{fig:turbine_solution} for a couple of snapshots of the von Mises stress distribution and blade deformation corresponding to two different pressure loadings. 

The suction-side $x_1$ and pressure-side $x_2$ pressures are varied in the following ranges: $x_1 \in [201160, 698390],~ x_2 \in [301460, 598020]$. A total of $n=30$ solution snapshots are generated, by sampling the inputs uniformly at random in $\mcl{X}\times \mcl{S}$, and supplied as seed points for the multifidelity run. We do the same for the HF case to keep the comparison fair. The \mfcv, HF, and random strategy experiments are then run for 50 iterations and are repeated 5 times with randomized seed points. The $q=2$ and $q=4$ experiments are run for 25 and 13 iterations, respectively.

\begin{figure}[htb!]
    \centering
    \begin{subfigure}{0.5\textwidth}
        \centering        \includegraphics[width=1\linewidth]{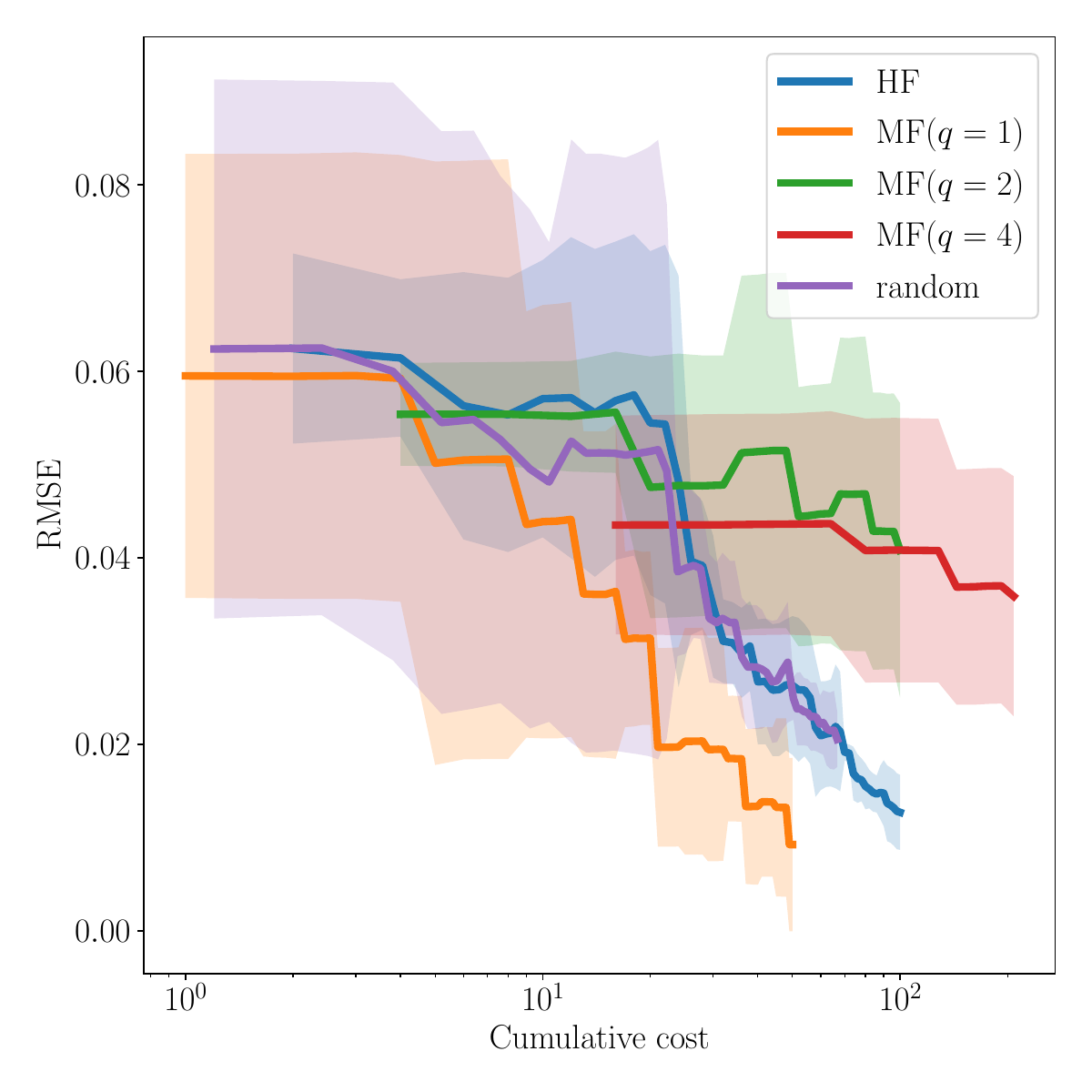}        
        \caption{RMSE versus cumulative cost.}
        \label{fig:turbine_mse}
    \end{subfigure}%
    \begin{subfigure}{0.5\textwidth}
        \centering        \includegraphics[width=1\linewidth]{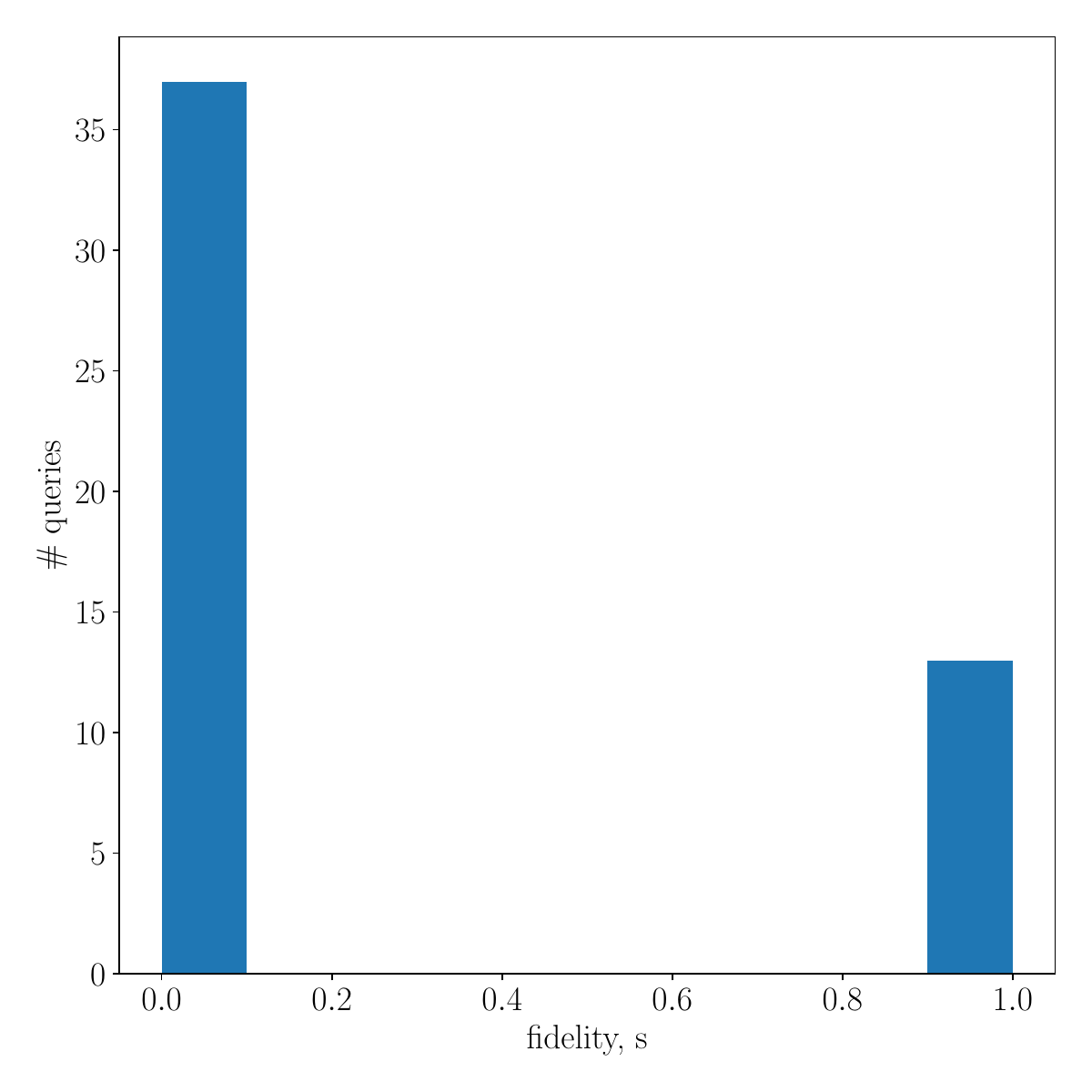}        
        \caption{Fidelity levels selected}
        \label{fig:turbine_fidselection}
    \end{subfigure}
    \caption{Turbine thermal stress experiment.}
    \label{fig:turbine}
\end{figure}

The plots of the RMSE versus cumulative cost is shown in \Cref{fig:turbine_mse}. Notice that \mfcv~with sequential acquisition ($q=1$) outperforms all other strategies. Batch sampling ($q=2$ and $q=4$) perform worse but, as in all other experiments, one of the variants of \mfcv~wins. Interestingly, the random sampling performs on par with the single-fidelity (HF). This is because, given that this experiment has only three discrete levels of fidelity, the random sampling occasionally samples new acquisitions at $s=1$. In the continuous fidelity case, the probability of sampling a point at $s=1$ is $0$; however, in this discrete fidelity case it is $0.33$, and hence this is expected. The fidelity levels automatically chosen by the algorithm ($q=1$) is shown in \Cref{fig:turbine_fidselection}. Notice that the high-fidelity model is queried approximately $25\%$ of the times, demonstrating the cost saving.


\section{Conclusion}
\label{sec:conclusion}
We present a novel active learning strategy to adaptively learn multifidelity Gaussian process models leveraging the leave-one-out cross-validation (LOO-CV) criterion. Our approach depends on maintainting, and continuously updating, two GPs. An ``outer'' GP that learns the mapping between $(\x,s)$ and models at various fidelities ($\hat{f}(\x,s)$). An an ``inner'' GP
that learns the logarithm of LOO-CV in the joint input-fidelity space ($\X\times \Sc$). Then, we use a two-step lookahead strategy to 
develop a multifidelity acquisition policy that allows both sequential as well as batch sampling. Our approach seamlessly works for both continuous and discrete fidelity spaces.

One of the major limitations of our work is that, as $n$ increases, computing \Cref{eqn:loocv_mv} for the LOO GP becomes intensive. Particularly, in the second line in \eqref{eqn:loocv_mv}, $n$ conjugate gradient steps are necessary. However, for the chosen experiments in this work, this was never a problem. Furthermore, we address situations where the high-fidelity model is very expensive to evaluate and only a few evaluations are affordable. In such situations, $n$ is not likely to be too high.
However, we believe it might be worth investigating efficient ways for computing the conjugate gradient steps, e.g., via parallelization and symbolic computations. This would be an avenue for futher work.
\clearpage
\section*{Appendix}
\label{sec:appendix}
Previously \citet{renganathan2023camera} and \citet{renganathan2022multifidelity} introduced several new multifideity test functions that can be used to benchmark multifidelity methods. We use a few of them, namely the 2D multifidelity multimodal function~\cite{bichon2008efficient}, the 2D multifidelity Four branches function~\cite{schobi2017rare}, and the 3D multifidelity Ishigami function~\cite{ishigami1990importance}. Additionally, we also use the 6D multifidelity Hartmann function provided in~\cite{simulationlib}. The specifics of these test functions are discussed as follows.

\subsubsection{2D multimodal function}
\begin{figure}[htb!]
    \centering
    \includegraphics[width=1\textwidth]{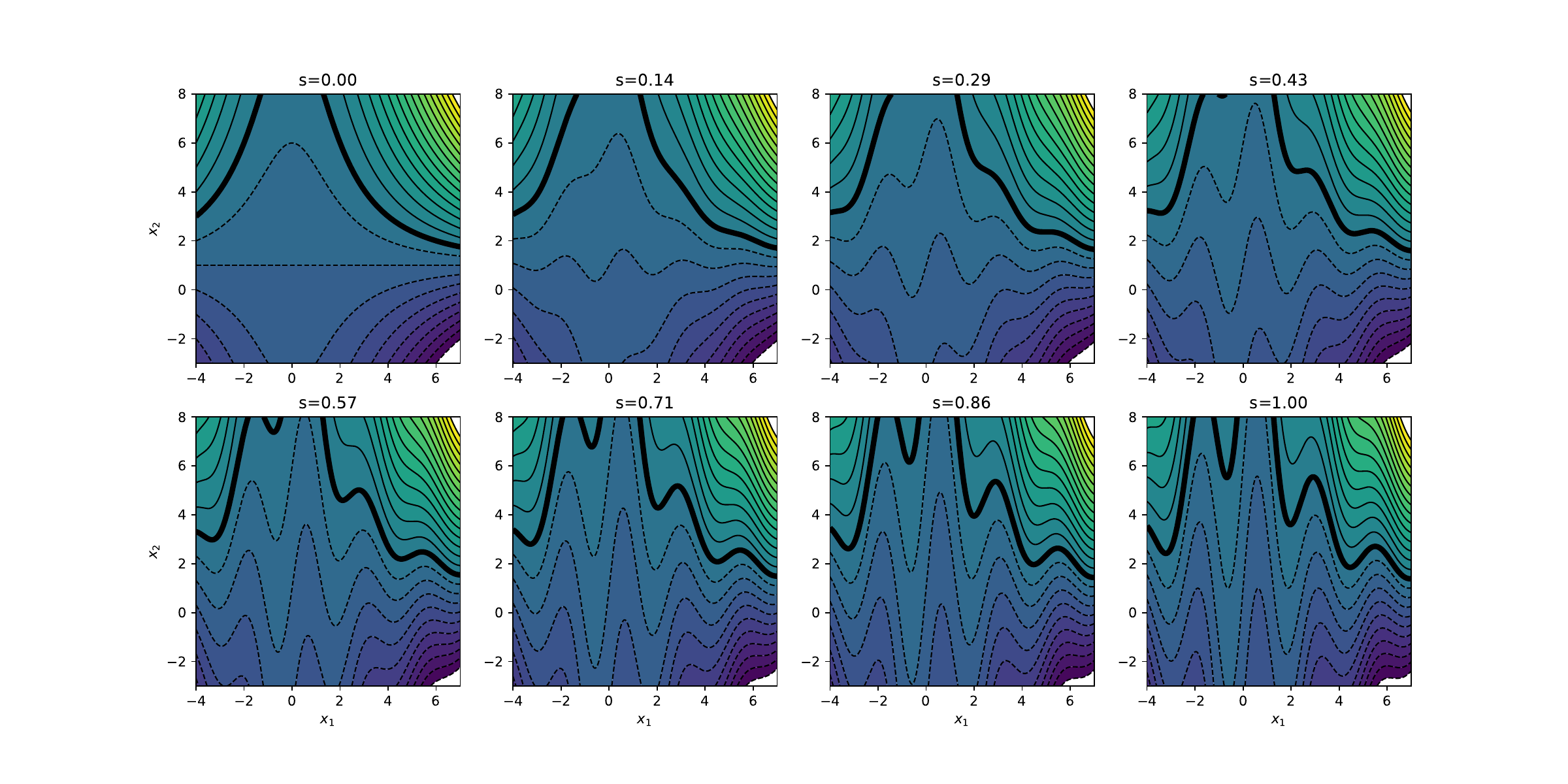}
    \caption{Multifidelity multimodal function from \cite{renganathan2023camera} }
    \label{f:mftestfunc1}
\end{figure}

We extend the test function in \cite{bichon2008efficient} to multifidelity setting as shown in \eqref{e:mf_multimodal}.

\begin{equation}
   f(\x, s) = \f{(x_1^2 + 4)(x_2 - 1)}{20} - s \times sin \left( \f{5x_1}{2} \right) - 2,
   \label{e:mf_multimodal}
\end{equation}
where $\x=[x_1, x_2]$, the domain $\X = [-4,7] \times [-3,8]$, and the last term introduces a fidelity dependence. This simple extension results in a correlated fidelity dependence, which is continuously differentiable in $\mcl{X}\times \mcl{S}$; see \Cref{f:mftestfunc1} for snapshots.

\subsubsection{2D Multifidelity Four-branches function}
\begin{figure}[htb!]
    \centering
    \includegraphics[width=1\textwidth]{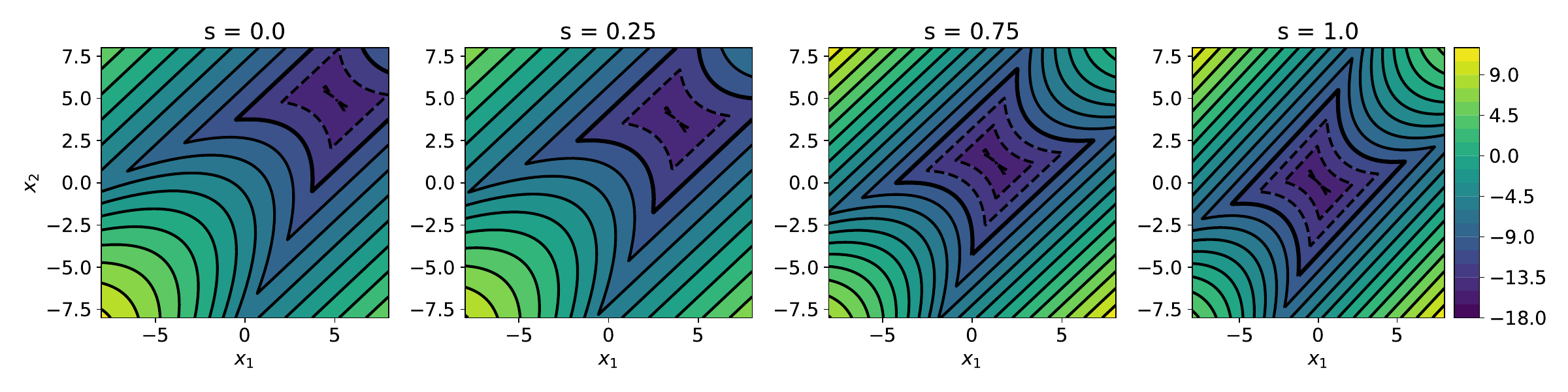}
    \caption{Multifidelity Four branches test function~\cite{renganathan2023camera}. }
    \label{f:mf_fourbranches}
\end{figure}

A system with four distinct component limit-states \cite{schobi2017rare} is a commonly used test problem in reliability analysis to demonstrate the efficacy of the algorithms in presence of multiple failure regions. The response function can be written as:
\begin{align}
    f({\x}, s) = \textrm{min} \begin{cases} 3 + 0.1(x'_1 -x'_2)^2 - \frac{x'_1 + x'_2}{\sqrt{2}}\,, \\
    3 + 0.1(x'_1 - x'_2)^2 + \frac{x'_1 + x'_2}{\sqrt{2}}\,, \\
    x'_1 - x'_2 + \frac{7}{\sqrt{2}}\,, \\
    x'_2 - x'_1 + \frac{7}{\sqrt{2}}\, 
    \end{cases}
\end{align}
where $\x = [x_1, x_2]$ and $\X = [-8,8]^2$. The multifidelity extension we provide, offers a translation of the level sets across fidelities, $x'_i = x_i - 5.0 \times s,~i=1,\ldots,3$, and is shown in \Cref{f:mf_fourbranches}.

\subsubsection{3D Multifidelity Ishigami function}
\begin{figure}
    \centering
    \includegraphics[width=1\textwidth]{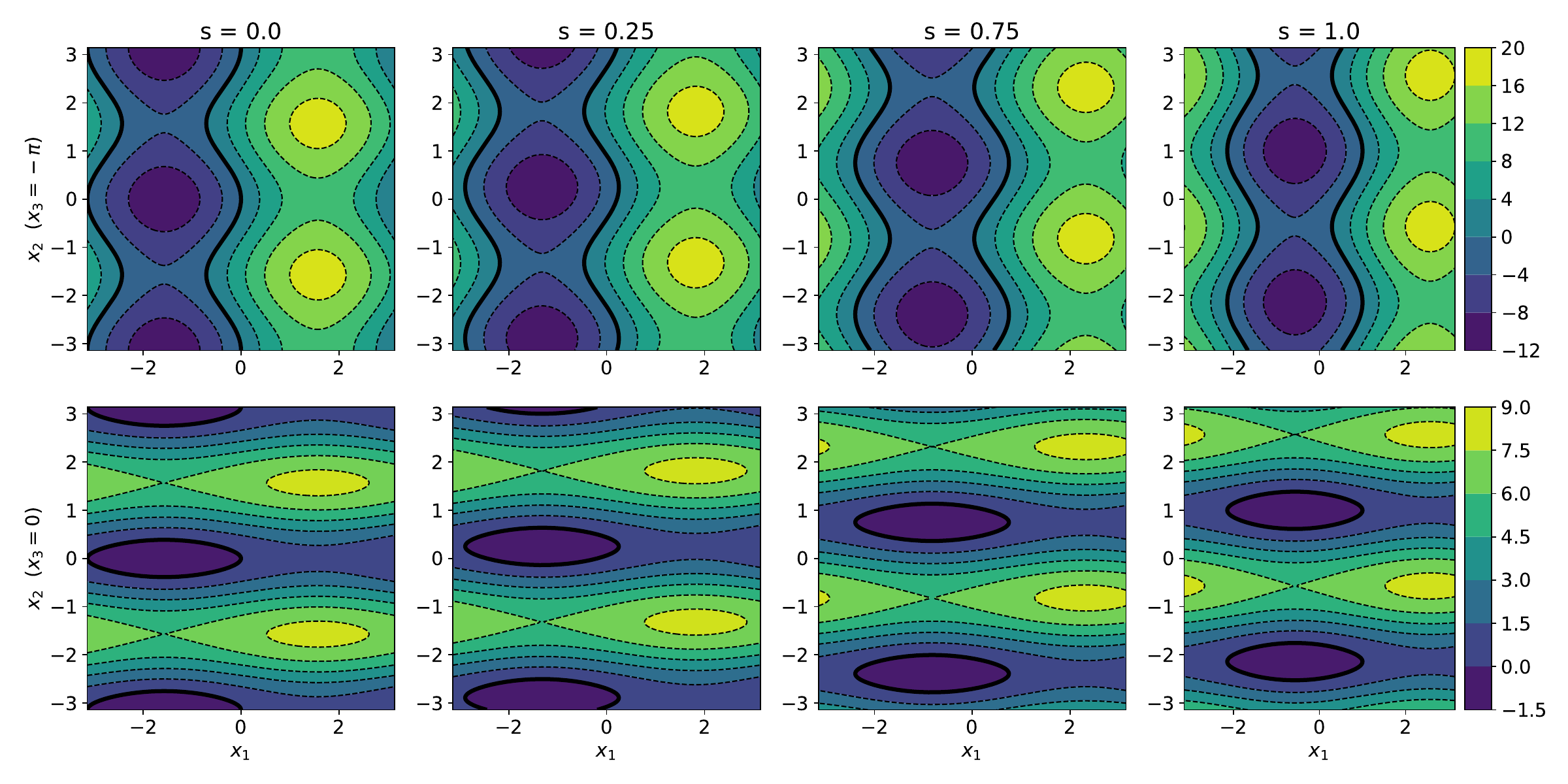}
    \caption{Multifidelity Ishigami test function~\cite{renganathan2023camera}. }
    \label{f:mf_ishigami}
\end{figure}

We modify the 3D Ishigami test function \cite{ishigami1990importance}, similar to the Four branches function, by introducing a shift in the variables using the fidelity parameter $s$, as shown in \eqref{e:mf_ishigami}. Note that, the input $\x$ is normalized to be in $[0,1]^3$ before introducing the translation. 
\begin{equation}
    f(\x, s) = \sin(x_1 - s) + 7.0 \sin^2(x_2 - s) + 0.1 x_3 ^4  \sin(x_1 - s),
    \label{e:mf_ishigami}
\end{equation}
where $\x = [x_1, x_2, x_3]$ and $\X=[-\pi, \pi]^3$. The resulting function is shown for two slices $x_3=-\pi$ and $x_3=0$ in \Cref{f:mf_ishigami}.

\subsubsection{6D Hartmann function}
\label{ss:6d_hartmann}
We use the "augmented" Hartmann 6D function introduced in \cite{balandat2019botorch} and restated below.
\begin{equation}
    f(\x, s) = -\left(\beta_1 - 0.1 * (1-s)\right) * \exp \left(- \sum_{j=1}^6 A_{1j} (x_j - P_{1j})^2 \right) -
            \sum_{i=2}^4 \beta_i \exp\left( - \sum_{j=1}^6 A_{ij} (x_j - P_{ij})^ 2\right),
    \label{e:mf_hartmann}
\end{equation}
where $\x = [x_1, \ldots, x_6]$ and $\X = [0,1]^6$ and
\[ 
A = \begin{bmatrix}
10 & 3 & 17 & 3.5 & 1.7 & 8 \\
0.05 & 10 & 17 & 0.1 & 8 & 14 \\
3 & 3.5 & 1.7 & 10 & 17 & 8 \\
17 & 8 & 0.05 & 10 & 0.1 & 14
\end{bmatrix}
\]
and
\[ 
P = \begin{bmatrix}
1312 & 1696 & 5569 & 124  & 8283 & 5886 \\
2329 & 4135 & 8307 & 3736 & 1004 & 9991 \\
2348 & 1451 & 3522 & 2883 & 3047 & 6650 \\
4047 & 8828 & 8732 & 5743 & 1091 & 381
\end{bmatrix}.
\]

\section*{Acknowledgments}
This work is supported by the Faculty Startup Funds at the The Pennsylvania State University. All computations were performed in the Penn State Institute of Computational and Data Sciences CPU cluster Roar.

\bibliographystyle{plainnat}
\bibliography{sample, ref, other_refs, cv}

\end{document}